\documentclass[12pt, draftclsnofoot, onecolumn]{IEEEtran}

\usepackage{cite}
\usepackage{algorithmic}
\usepackage{array}
\usepackage{bbm}
\usepackage{amsfonts}
\usepackage{graphicx}
\usepackage[hidelinks]{hyperref}
\usepackage{xcolor}
\hypersetup{
	colorlinks  = true,
	linkcolor	={red!100!black},
	citecolor	={blue!100!black},
	urlcolor	={blue!100!black}
}
\usepackage{mathtools}
\usepackage{subfigure,epsfig}
\usepackage{amsmath, amsthm, amssymb}
\usepackage[outdir=./]{epstopdf}
\usepackage{dsfont}
\usepackage{mathrsfs}
\newtheorem{theorem}{Theorem}
\newtheorem{Corollary}{Corollary}
\newtheorem{Definition}{Definition}

\def\figref#1{Fig.\,\ref{#1}}%
\newlength{\figwidth}
\begin{document}
\title{Accuracy of Distance-Based Ranking of Users in the Analysis of NOMA Systems}
\author{Mohammad Salehi, Hina Tabassum, and Ekram Hossain\thanks{M. Salehi and E. Hossain are with the Department of Electrical and Computer Engineering at the University of Manitoba, Canada. H. Tabassum is with the Department of Electrical Engineering and Computer Science at York University, Canada 
(emails: salehim@myumanitoba.ca, hina@eecs.yorku.ca, Ekram.Hossain@umanitoba.ca). \textbf{E. Hossain} is the corresponding author.
 }}

\maketitle

\begin{abstract}	
We characterize the accuracy of analyzing the performance of a non-orthogonal multiple access (NOMA)  system where users are ranked according to their distances instead of instantaneous channel gains, i.e.,  product of their distance-based path-loss and fading channel gains. Distance-based ranking of users is analytically tractable and can lead to important insights. However, it may not be appropriate in a multipath fading environment where a near user suffers from severe fading while a far user experiences weak fading. Since the ranking of users (and in turn interferers) in a NOMA system  has a direct impact on coverage probability analysis, impact of the traditional distance-based ranking, as opposed to instantaneous signal power-based ranking, needs to be understood. This will enable us to identify scenarios where distance-based ranking, which is easier to implement compared to instantaneous signal power-based ranking, is acceptable for system performance analysis. To this end, in this paper, we derive  the probability of the event when distance-based ranking yields the same results as instantaneous signal power-based ranking, which is  referred to as the {\em accuracy probability}.  We characterize the probability of accuracy considering Nakagami-$m$ fading channels and three different spatial distribution models of user locations in NOMA, namely, Poisson Point Process (PPP), Matern Cluster Process (MCP), and Thomas Cluster Process (TCP). For all  these models of users' locations, we assume that the spatial locations of the base stations (BSs) follow a homogeneous PPP.  We show that the accuracy probability decreases with the increasing number of users and increases with the path-loss exponent. In addition, through examples, we illustrate  the impact of accuracy probability on uplink and downlink coverage probability. Closed-form expressions are presented for Rayleigh fading environment.  Effects of fading severity and users' pairing on the accuracy probability are also investigated.
\end{abstract}

\newpage
\section{Introduction}

Performance of non-orthogonal multiple access (NOMA) in both uplink and downlink depends on the successive intra-cell interference cancellation (SIC) which relies on the  ranking of the users in each NOMA cluster \cite{2017uplink}. In particular, downlink intra-cell interference received at a given user in NOMA depends on the power allocation factors of users in the cluster. These power allocation factors are designed according to the ranking of users' transmission links quality. For example, users with stronger links have smaller power allocations and vice versa. On the other hand, in uplink NOMA, to apply SIC, BS successively decodes and cancels the messages of strong channel users, prior to decoding the signals of weak channel users \cite{2017uplink}. Therefore, the intra-cell interference encountered by any user depends on the instantaneous received signal powers (which includes short-term fading) of users in the NOMA cluster.

The link quality can be evaluated by different metrics. These metrics should include effects of path-loss (and therefore link distance), fading, and/or inter-cell interference\footnote{Note that uplink inter-cell interference at the desired BS is same for different users in a NOMA cluster.  Therefore, signal power-based ranking and  signal-to-intercell-interference-ratio (SINR)-based ranking yield the same result.}~\cite{ali2018downlink}.  However, acquiring complete channel state information (CSI) with fading and inter-cell interference increases system complexity. Therefore, most of the existing state-of-the-art  resorts to mean signal power- (or distance-) based user ranking in NOMA analysis. Recently, in \cite{Tabassum2017},   the rate coverage probability of a user at rank $m$ in uplink NOMA has been derived assuming distance-based ranking. In \cite{Wildemeersch2014,Geraci2016}, it is assumed that the order statistics of instantaneous signal power are dominated by the distance; hence, in the analysis, users are ordered based on their distances instead of complete CSI. In \cite{Shi2018,Choi2016}, distance-based ranking is used for the analysis of NOMA systems with HARQ. In \cite{Liu2016jsac}, the authors  study two-user cooperative NOMA and derive the outage probability  assuming the near user to be  the strong user and the far user to be  the weak user. In \cite{Ding2016twc}, a similar assumption is made for the analysis of uplink and downlink MIMO NOMA. In order to maximize the rate region of the uplink NOMA systems, in \cite{Chingoska2016letter}, decoding order of the information signal {at the BS} is the inverse of the distances. In \cite{ali2018downlink}, the authors derive the outage probability in downlink Poisson cellular networks where users are ranked based on mean signal power and {instantaneous SINR}. 

To avoid analytical complexity (in theory) and  overcome implementation complexity, mean signal power- (distance-) based ranking is typically considered to be appropriate for ordering users in a NOMA cluster.  Although this method simplifies the analysis and provides tractable results, its validation (i.e., {\em accuracy}) has not been studied yet. The distance-based ordering may not always be accurate, especially in a dynamic multipath fading environment, where a near user can experience severe fading and a far user  can observe weak fading.
Since the ranking of users in a NOMA system  has a direct impact on the system performance (e.g., coverage probability) analysis, 
it is crucial to quantify the impact of distance-based ranking  in various environments and to identify the scenarios where this ranking is accurate (i.e., provides system performance close to that achievable with full CSI-based  user ranking). 

The contributions of this paper can be summarized as follows:
\begin{itemize}
\item This paper characterizes the accuracy of analyzing the performance of a NOMA system where users are ranked according to their distances (or, equivalently, mean signal powers) instead of instantaneous signal powers, i.e.,  product of their distance-based path-loss  and fading channel. In particular, we derive  the probability of the event when distance-based ranking yields same results as instantaneous signal power-based ranking, which is referred to as  the {\em accuracy probability}. 

\item We characterize the accuracy probability  considering Nakagami-$m$ fading and three different spatial distribution models of user locations in NOMA, namely, Poisson Point Process (PPP), Matern Cluster Process (MCP), and Thomas Cluster Process (TCP). For all three user location models, the  spatial locations of the BSs are assumed to follow a homogeneous PPP. The expressions are applicable to both uplink and downlink NOMA scenarios. 

\item By analyzing the properties of the derived accuracy probability, we show that the accuracy probability decreases with the increasing number of NOMA users and increases with the path-loss exponent. Closed-form expressions are derived for special cases with two and three users in a NOMA cluster and Rayleigh fading. In addition, through examples, we illustrate  the impact of accuracy probability on uplink and downlink. We observe that the impact of distance-based ranking on network performance metrics such as coverage probability  is different in the uplink and the downlink.

\item Using the derived expressions, we obtain following insights: (i) For the PPP model, the accuracy probability does not depend on BS intensity $\lambda$, (ii) For the MCP model, the accuracy probability does not depend on cluster radius $R$, (iii) For TCP model, the accuracy probability does not depend on scattering variance $\sigma^2$, which is a measure of cluster size.

\item Finally, we study the impact of fading severity and user selection on the accuracy probability. 
\end{itemize}

The rest of the paper is organized as follows. The system model and assumptions are presented in Section~II. The definition and properties of the accuracy probability are provided in Section III along with the discussions on their impact on uplink and downlink coverage probability. For Rayleigh and Nakagami-$m$ fading, the accuracy probability is derived in Sections IV and V, respectively. In Section VI, the impact of user pairing on the accuracy probability is investigated. Numerical results are presented in Section VII. Finally, Section VIII concludes the paper.

\section{System Model and Assumptions}
We assume that the spatial locations of the BSs follow a homogeneous Poisson point process (PPP) $\Phi$ of intensity $\lambda$ and those of the users follow three different models as described in the following.
\begin{itemize}
\item {\bf PPP:} Users are distributed according to a homogeneous PPP $\Phi_{\rm U}$ of intensity $\lambda_{\rm u}$ and each user is associated to its nearest BS. We consider a heavily loaded regime, i.e., $\lambda_{\rm u}\gg\lambda$ where we have at least $N$ users in a typical Voronoi cell\footnote{In a heavily loaded network, when $N$ is small, assuming that we have at least $N$ users in a typical cell, is not unrealistic. For instance, when $\lambda_{\rm u}/\lambda=10$, according to \cite{Yu2013}[Lemma 1], the probability of having more than one user in the typical cell is 0.97 and probability of having more than two users is 0.93.}. 
To form a NOMA cluster of size $N$ in the typical cell, we randomly select $N$ users. Therefore, NOMA users are uniformly distributed within the typical Voronoi cell. The explicit distribution of the main geometrical characteristics of the typical cell of a Voronoi tessellation is not known \cite{Renzo2018}. In \cite{Mukherjee2012,Haenggi2017user,Wang2017}, taking $c=5/4$, the probability density function (PDF) and the cumulative distribution function (CDF) of the distance for a typical user from its serving BS can  be approximated, respectively, as follows:
\begin{IEEEeqnarray}{rCl}
	f_r(x)\approx 2 c \lambda \pi x e^{-c \lambda \pi x^2}, \qquad
	F_r(x)\approx 1 -               e^{-c \lambda \pi x^2}, \qquad   x\ge 0.
	\label{eq:PDF-PPP}
\end{IEEEeqnarray} 
\item {\bf Matern Cluster Process (MCP):} Users are spatially distributed according to an MCP, where the BS point process $\Phi$ is the parent point process. In each NOMA cluster, $N$ users are uniformly distributed in a ball of radius $R$ centered at the serving BS\footnote{For MCP and TCP models, we also assume that network is heavily loaded.}. The PDF and CDF of the link distance from an arbitrary user in a cluster {to its serving BS} are given, respectively, as:
\begin{IEEEeqnarray}{rCl}
	f_r(x)=\frac{2x}{R^2}\mathbf{1}(0 \le x \le R), \qquad
	F_r(x)=\frac{x^2}{R^2}\mathbf{1}(0 \le x \le R),
	\label{eq:PDF-MCP}
\end{IEEEeqnarray} 
where $\mathbf{1}(.)$ is the indicator function.

\item {\bf Thomas Cluster Process (TCP):}
Users are distributed according to a TCP, where BS point process $\Phi$ is the parent point process. Each NOMA cluster is formed by randomly selecting $N$ users from the set of users that have the same parent. The PDF and the CDF of the link distance between an arbitrary user and its serving BS are given, respectively, as follows:
\begin{IEEEeqnarray}{rCl}
	f_r(x)=\frac{x}{\sigma^2} \exp \left\{-\frac{x^2}{2\sigma^2}\right\}, \qquad 
	F_r(x)=1     -            \exp \left\{-\frac{x^2}{2\sigma^2}\right\}, \qquad x\ge 0.
	\label{eq:PDF-TCP}
\end{IEEEeqnarray} 
In particular, $N$ users are independently and identically distributed following a normal distribution with variance $\sigma^2$ around each BS. 
\end{itemize}
Let us denote the distance between the $i$-th nearest user (termed rank $i$ user) and the serving BS by $r_{(i)}$, $1\le i\le N$. The received power, for the user at rank $i$, is modeled by $h_ir_{(i)}^{-\alpha}$. $r_{(i)}^{-\alpha}$ represents the large-scale path-loss where $\alpha>2$ is the path-loss exponent. $h_i$ models the channel power gain due to small-scale fading. The channel power gains follow independent gamma distribution with parameter $m$ and mean $\Omega$ for Nakagami-$m$ fading environment, i.e.,
\begin{IEEEeqnarray}{rCl}
	f_h(x)=\frac{m^m x^{m-1}}{\Gamma(m) \Omega^m}\exp\left( -\frac{mx}{\Omega} \right),
\end{IEEEeqnarray} 
where $\Gamma(.)$ is the gamma function. By setting $m=1$, it reduces to the exponential distribution, corresponding to Rayleigh fading.

\section{Probability of Accuracy: Definition and Properties}
Ranking users based on their distances from the serving BS in each NOMA cluster is a common assumption in the existing literature to characterize the performance of NOMA. That is, the nearest user to the serving BS is assumed as the user with the highest CSI and so on (which may not always be true). To understand the accuracy of this approximation and its impact on important performance metrics such as coverage probability, in this section, we define the term {\em accuracy probability} $\mathcal{A}$, highlight its properties, and describe its connection to uplink and downlink coverage probability through examples. 

\begin{Definition}[Accuracy probability] \label{Def1}
Accuracy probability $\mathcal{A}$ is the probability that ordering based on large-scale path loss\footnote{For a fixed path-loss exponent, we use path-loss-based ranking and  ``distance-based ranking'' interchangeably throughout the paper.} matches ordering based on the instantaneous signal power (small-scale fading and large-scale path-loss), i.e.,
	\begin{IEEEeqnarray}{rCl}
		\mathcal{A}=\mathbb{P}\left( h_1r_{(1)}^{-\alpha}>h_2r_{(2)}^{-\alpha}>\cdots>h_Nr_{(N)}^{-\alpha} \right).
		\label{eq:match-probability}
	\end{IEEEeqnarray}
    Therefore, ordering users based on path-loss, instead of instantaneous signal power is accurate with probability $\mathcal{A}$.
\end{Definition}
Using the indicator function, the accuracy probability can be expressed as
\begin{IEEEeqnarray}{rCl}
	\mathcal{A}&=&\mathbb{E}_{\{h_i\},\{r_{(i)}\}}\left[ \mathbf{1} 
	\left( h_1r_{(1)}^{-\alpha}>h_2r_{(2)}^{-\alpha}>\cdots>h_Nr_{(N)}^{-\alpha} \right) \right] 
	\nonumber \\
	&=& \mathbb{E}_{\{r_{(i)}\}} \left[ \mathbb{E}_{\{h_i\}}\left[ \mathbf{1} 
	\left( h_1r_{(1)}^{-\alpha}>h_2r_{(2)}^{-\alpha}>\cdots>h_Nr_{(N)}^{-\alpha} \right) \right] \right].
	\label{eq:match-expectation}
\end{IEEEeqnarray}
The inner expectation in \eqref{eq:match-expectation} is over the channel power gains $\left\{ h_i \right\}$, i.e., the inner expectation calculates the accuracy probability for a given realization of users and BSs. The outer expectation is with respect to the ordered desired link distances $\left\{ r_{(i)} \right\}$. In the derivation of the outer expectation we use the following definition.

\begin{figure}
	\parbox[c]{.5\textwidth}{%
		\centerline{\subfigure[$\beta=0$ (perfect SIC).]
			{\epsfig{file=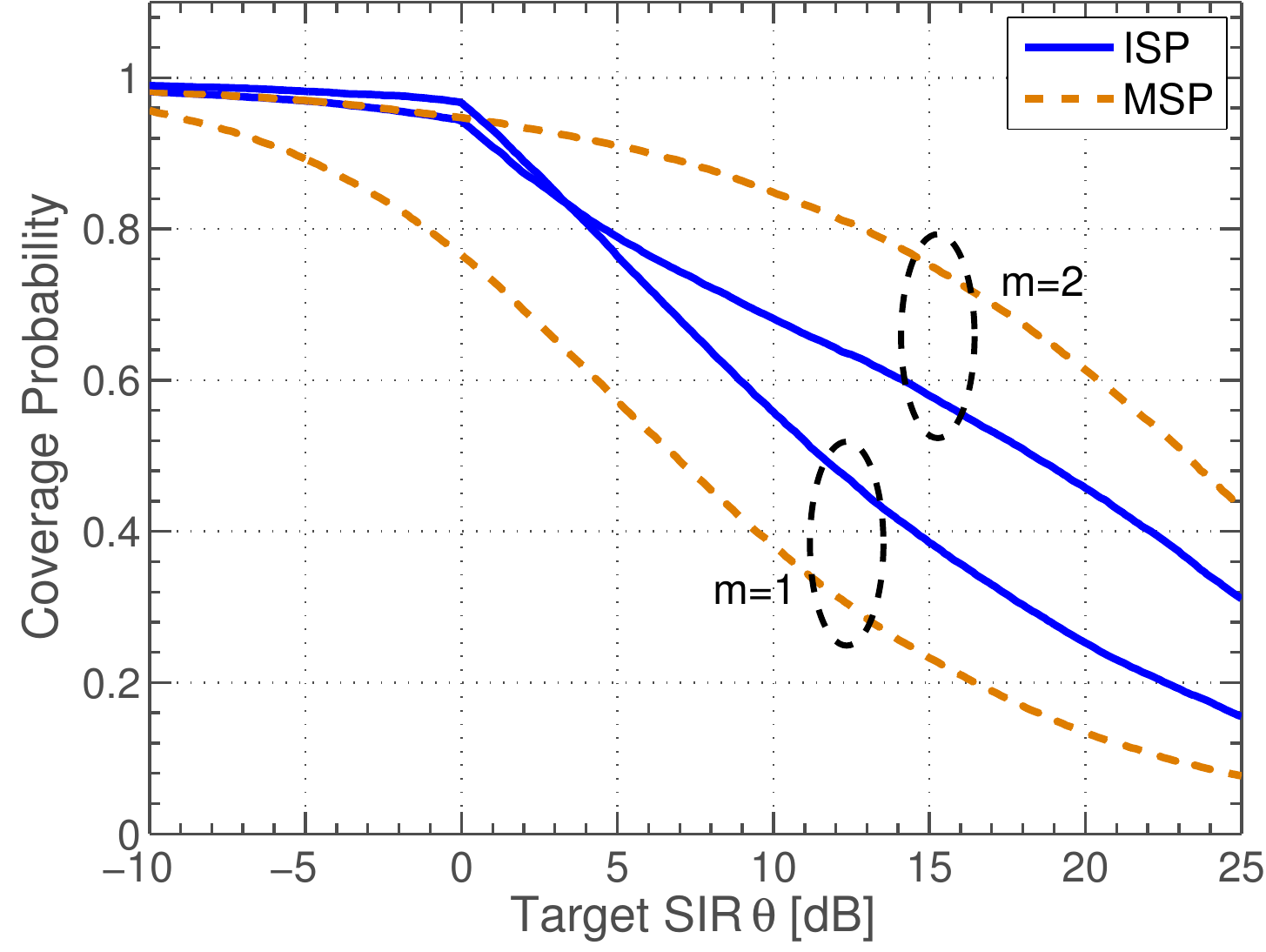,width=.49\textwidth,height=0.9\figwidth}}}} 
	\parbox[c]{.5\textwidth}{%
		\centerline{\subfigure[$\beta=0.5$.]
			{\epsfig{file=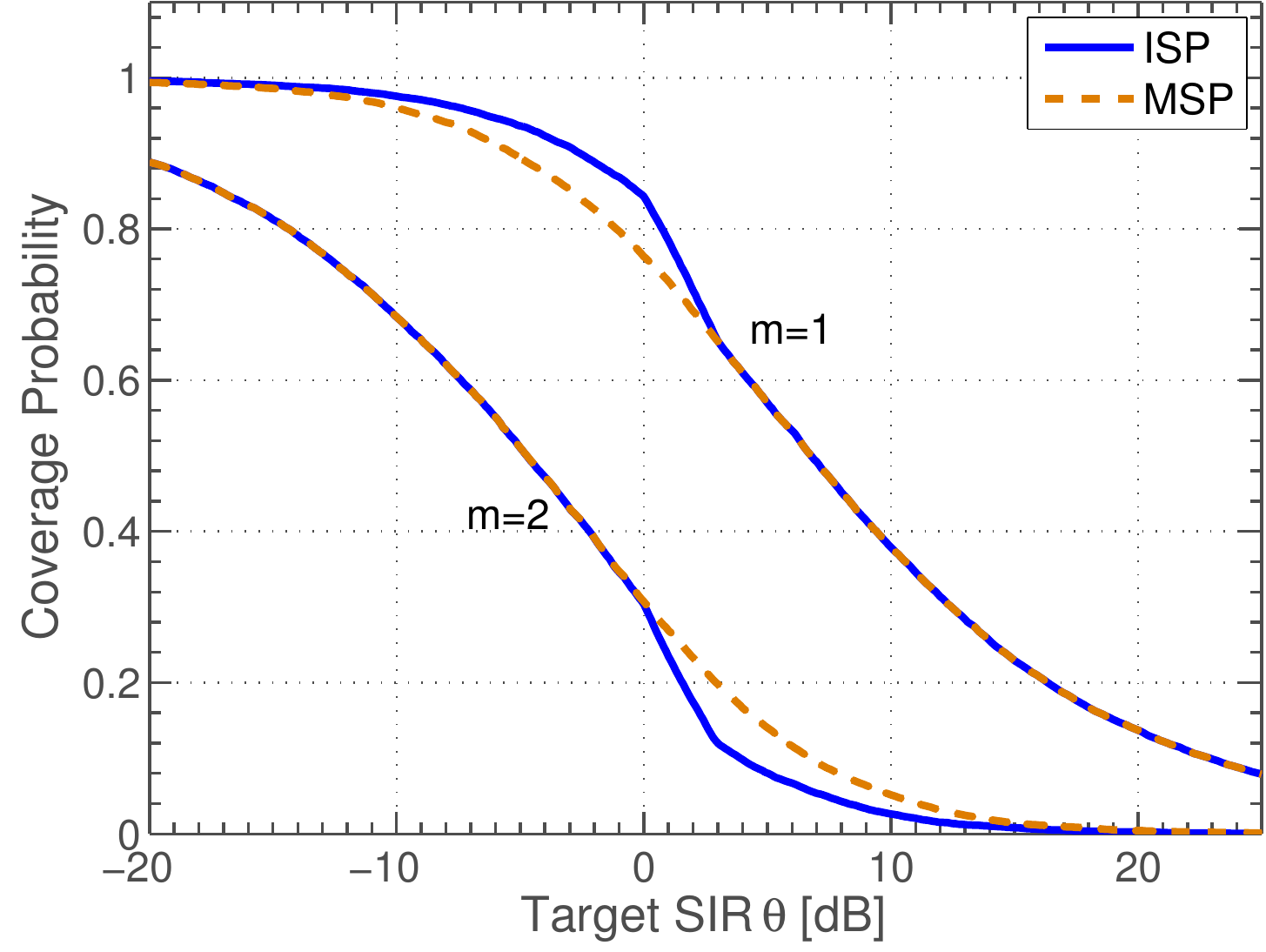,width=.49\textwidth,height=0.9\figwidth}}}} 
	\caption{Uplink coverage probability for MCP model with $N=2$. $\lambda=0.0001$, $R=10$, $\sigma_n^2=0$, and $\alpha=4$. Coverage probabilities that are derived based on \eqref{eq:example1_1} and \eqref{eq:example1_2} are called ``ISP'' in the legend, and their approximations which are obtained by assuming $\mathbb{P}\left( h_1 r_{(1)}^{-\alpha}>h_2 r_{(2)}^{-\alpha} \right) \approx 1$ are called ``MSP''.}
	\label{fig:Coverage_uplink}
\end{figure}

\begin{Definition}[Joint PDF of $N$-ordered Random Variables]
Let $r_1,r_2,...,r_N$ be a set of $N$ i.i.d. random variables with PDF $f_r(x)$. Let $r_{(i)}$ denote the $i$-th smallest observation of the $N$ random variables, i.e., $r_{(1)} \le r_{(2)} \le \cdots \le r_{(N)}$. The joint PDF of $N$-ordered random variables can then be given as \cite{yang2011order}:
\begin{IEEEeqnarray}{rCl}
	f_{r_{(1)},r_{(2)},...,r_{(N)}}(x_1,x_2,...,x_N)=N! \prod_{i=1}^{N}f_r(x_i), \qquad   x_1\le x_2\le \cdots \le x_N. 
	\label{eq:statistic}
\end{IEEEeqnarray}
\end{Definition} 
In the following, two properties of the accuracy probability are reported. These properties are general and apply to any of the considered fading channel and users' spatial distributions.
\begin{Corollary}
The accuracy probability $\mathcal{A}$ fulfills the following properties: i) $\mathcal{A}$ is a decreasing function of NOMA cluster size $N$, ii)  $\mathcal{A}$ is an increasing function of path-loss exponent $\alpha$. 
\end{Corollary}
\begin{IEEEproof}
The result in (i)  follows from definition of the accuracy probability \eqref{eq:match-probability}. The result in (ii) follows from \eqref{eq:match-expectation}. If $h_1r_{(1)}^{-\alpha}>h_2r_{(2)}^{-\alpha}>\cdots>h_Nr_{(N)}^{-\alpha}$ is satisfied by $\alpha$, it will also be satisfied by higher values of path-loss exponent. On the other hand, if $h_1r_{(1)}^{-\alpha}>h_2r_{(2)}^{-\alpha}>\cdots>h_Nr_{(N)}^{-\alpha}$ is not satisfied by $\alpha$, any smaller value of path-loss exponent cannot also satisfy this condition. Therefore, when we increase the path-loss exponent, ranking users based on their distances is valid for wider range of channel and distance realizations, i.e., $\mathcal{A}$ is an increasing function of  $\alpha$.
\end{IEEEproof}
Now we discuss the impact of ranking method  on uplink and downlink coverage probability, respectively, as follows:

\noindent
{{\bf Example  - Uplink NOMA:}  To apply SIC, in each step, BS decodes the signal of user with the highest instantaneous signal power by treating other signals as noise. Therefore, for the 2-UE NOMA, the coverage probability of the near user to the BS ($ P_{{\rm cov},(1)}^{\rm ISP}$) should be derived as follows:
\begin{IEEEeqnarray}{rCl}
    P_{{\rm cov},(1)}^{\rm ISP}&=&\mathbb{P} \left\{ 
    \frac{P_{\rm tx}h_1 r_{(1)}^{-\alpha}}{P_{\rm tx}h_2 r_{(2)}^{-\alpha} + I_{\rm inter} + \sigma_n^2} > \theta \mid h_1 r_{(1)}^{-\alpha}>h_2 r_{(2)}^{-\alpha} \right\} \underbrace{\mathbb{P}\left( h_1 r_{(1)}^{-\alpha}>h_2 r_{(2)}^{-\alpha} \right)}_{\mathcal{A}} \nonumber 
    \\
    && +\> \mathbb{P} \left\{ \frac{P_{\rm tx}h_1 r_{(1)}^{-\alpha}}{\beta P_{\rm tx}h_2 r_{(2)}^{-\alpha} + I_{\rm inter} + \sigma_n^2} > \theta \mid h_1 r_{(1)}^{-\alpha}<h_2 r_{(2)}^{-\alpha} \right\} \mathbb{P}\left( h_1 r_{(1)}^{-\alpha}<h_2 r_{(2)}^{-\alpha} \right),  
    \label{eq:example1_1}
\end{IEEEeqnarray}
where  $I_{\rm inter}$ denotes the inter-cell interference; $\sigma_n^2$ is the noise power, $P_{\rm tx}$ is the transmit power, and $\beta\in[0,1]$ captures the effect of imperfect SIC. According to \eqref{eq:example1_1}, when $h_1 r_{(1)}^{-\alpha}>h_2 r_{(2)}^{-\alpha}$, BS decodes the intended signal of near user in the presence of interference from far user, and when $h_1 r_{(1)}^{-\alpha}<h_2 r_{(2)}^{-\alpha}$, BS decodes and cancels the signal of far user and then decodes the intended signal of near user. Similarly, for the far user, we have  
\begin{IEEEeqnarray}{rCl}
	P_{{\rm cov},(2)}^{\rm ISP}&=&\mathbb{P} \left\{ 
	\frac{P_{\rm tx}h_2 r_{(2)}^{-\alpha}}{\beta P_{\rm tx}h_1 r_{(1)}^{-\alpha} + I_{\rm inter} + \sigma_n^2} > \theta \mid h_1 r_{(1)}^{-\alpha}>h_2 r_{(2)}^{-\alpha} \right\} {\mathbb{P}\left( h_1 r_{(1)}^{-\alpha}>h_2 r_{(2)}^{-\alpha} \right)} \nonumber 
	\\
	&& +\> \mathbb{P} \left\{ \frac{P_{\rm tx}h_2 r_{(2)}^{-\alpha}}{ P_{\rm tx}h_1 r_{(1)}^{-\alpha} + I_{\rm inter} + \sigma_n^2} > \theta  \mid h_1 r_{(1)}^{-\alpha}<h_2 r_{(2)}^{-\alpha} \right\} \mathbb{P}\left( h_1 r_{(1)}^{-\alpha}<h_2 r_{(2)}^{-\alpha} \right).  
	\label{eq:example1_2}
\end{IEEEeqnarray}}
{In the analysis of uplink NOMA, it is generally assumed that the nearest user to the BS has the highest instantaneous signal power, i.e., $\mathbb{P}\left( h_1 r_{(1)}^{-\alpha}>h_2 r_{(2)}^{-\alpha} \right) \approx 1$ for 2-UE NOMA.  Hence, only the first terms in \eqref{eq:example1_1} and \eqref{eq:example1_2} are derived to date in the literature and reported as $P_{{\rm cov},(1)}^{\rm ISP}$ and $P_{{\rm cov},(2)}^{\rm ISP}$, respectively. However, the first terms in \eqref{eq:example1_1} and \eqref{eq:example1_2} provide good approximations only when: i) network is intercell-interference- (or noise-) limited, ii) $\beta$ is close to 1, i.e, unsuccessful SIC is very likely, or iii) assumption $\mathbb{P}\left( h_1 r_{(1)}^{-\alpha}>h_2 r_{(2)}^{-\alpha} \right) \approx 1$ is accurate. In \figref{fig:Coverage_uplink}, $P_{{\rm cov},(1)}^{\rm ISP}$, $P_{{\rm cov},(2)}^{\rm ISP}$, and their approximations based on the assumption $\mathbb{P}\left( h_1 r_{(1)}^{-\alpha}>h_2 r_{(2)}^{-\alpha} \right) \approx 1$ (in the plots denoted by the legend ``MSP") are provided. According to \figref{fig:Coverage_uplink}, the coverage probability can be significantly different for the distance-based ranking and full CSI-based ranking.}

\noindent
{{\bf Example  - Downlink NOMA:} Order of decoding at a given user depends on the power allocations of users' signals at the BS. For 2-UE downlink NOMA, BS typically allocates more power to the weak user; thus, weak user can decode its intended signal in the presence of interference from the strong user. On the other hand, the strong user decodes and cancels the signal of weak user before decoding its intended signal. Therefore, with instantaneous signal power-based ranking at the BS, the coverage probabilities of near and far users are as follows:
\begin{IEEEeqnarray}{rCl}
	P_{{\rm cov},(1)}^{\rm ISP}&=&\mathbb{P} \left\{ 
	\frac{a_1 P_{\rm BS}h_1 r_{(1)}^{-\alpha}}{\beta a_2 P_{\rm BS}h_1 r_{(1)}^{-\alpha} + I_{\rm inter}^{(1)} + \sigma_n^2} > \theta \mid h_1 r_{(1)}^{-\alpha}>h_2 r_{(2)}^{-\alpha} \right\} \mathbb{P}\left( h_1 r_{(1)}^{-\alpha}>h_2 r_{(2)}^{-\alpha} \right) \nonumber 
	\\
	&& +\> \mathbb{P} \left\{ \frac{a_2 P_{\rm BS}h_1 r_{(1)}^{-\alpha}}{a_1 P_{\rm BS}h_1 r_{(1)}^{-\alpha} + I_{\rm inter}^{(1)} + \sigma_n^2} > \theta \mid h_1 r_{(1)}^{-\alpha}<h_2 r_{(2)}^{-\alpha} \right\} \mathbb{P}\left( h_1 r_{(1)}^{-\alpha}<h_2 r_{(2)}^{-\alpha} \right),  \quad
	\label{eq:example2_1}
\end{IEEEeqnarray}
\begin{IEEEeqnarray}{rCl}
	P_{{\rm cov},(2)}^{\rm ISP}&=&\mathbb{P} \left\{ 
	\frac{a_2 P_{\rm BS}h_2 r_{(2)}^{-\alpha}}{a_1 P_{\rm BS}h_2 r_{(2)}^{-\alpha} + I_{\rm inter}^{(2)} + \sigma_n^2} > \theta \mid h_1 r_{(1)}^{-\alpha}>h_2 r_{(2)}^{-\alpha} \right\} {\mathbb{P}\left( h_1 r_{(1)}^{-\alpha}>h_2 r_{(2)}^{-\alpha} \right)} \nonumber 
	\\
	&& +\> \mathbb{P} \left\{ \frac{a_1 P_{\rm BS}h_2 r_{(2)}^{-\alpha}}{ \beta a_2 P_{\rm BS}h_2 r_{(2)}^{-\alpha} + I_{\rm inter}^{(2)} + \sigma_n^2} > \theta  \mid h_1 r_{(1)}^{-\alpha}<h_2 r_{(2)}^{-\alpha} \right\} \mathbb{P}\left( h_1 r_{(1)}^{-\alpha}<h_2 r_{(2)}^{-\alpha} \right). \nonumber \\
	\label{eq:example2_2}
\end{IEEEeqnarray}
$a_1 P_{\rm BS}$ and $a_2 P_{\rm BS}$ denote the allocated powers to the strong and weak users where $0<a_1<a_2<1$ and $a_1+a_2=1$. Note that, unlike uplink, inter-cell interference seen at different users is different in downlink; $I_{\rm inter}^{(1)}$ and $I_{\rm inter}^{(2)}$ denote the inter-cell interference at the near and far users, respectively. With instantaneous signal power-based ranking at the BS, when $h_1 r_{(1)}^{-\alpha}>h_2 r_{(2)}^{-\alpha}$, BS allocates more power to the far (weak) user, while, when $h_1 r_{(1)}^{-\alpha}<h_2 r_{(2)}^{-\alpha}$, more power is allocated to the near (weak) user. On the other hand, with distance-based ranking at the BS, BS always allocates more power to the far user, i.e., far user is always considered as the weak user. Therefore, coverage probabilities of the near and far users with distance-based ranking at the BS can be derived by the first terms in \eqref{eq:example2_1} and \eqref{eq:example2_2}, respectively. In \figref{fig:Coverage_downlink}, we compare the coverage probabilities for distance-based and instantaneous signal power-based ranking. Note that coverage probabilities with distance-based ranking provide close results to the coverage probabilities with instantaneous signal power-based ranking when: i) the network is intercell-interference- (or noise-) limited, or ii) the assumption $\mathbb{P}\left( h_1 r_{(1)}^{-\alpha}>h_2 r_{(2)}^{-\alpha} \right) \approx 1$ is accurate. 
\begin{figure}
	\parbox[c]{.5\textwidth}{%
		\centerline{\subfigure[$\beta=0$ (perfect SIC).]
			{\epsfig{file=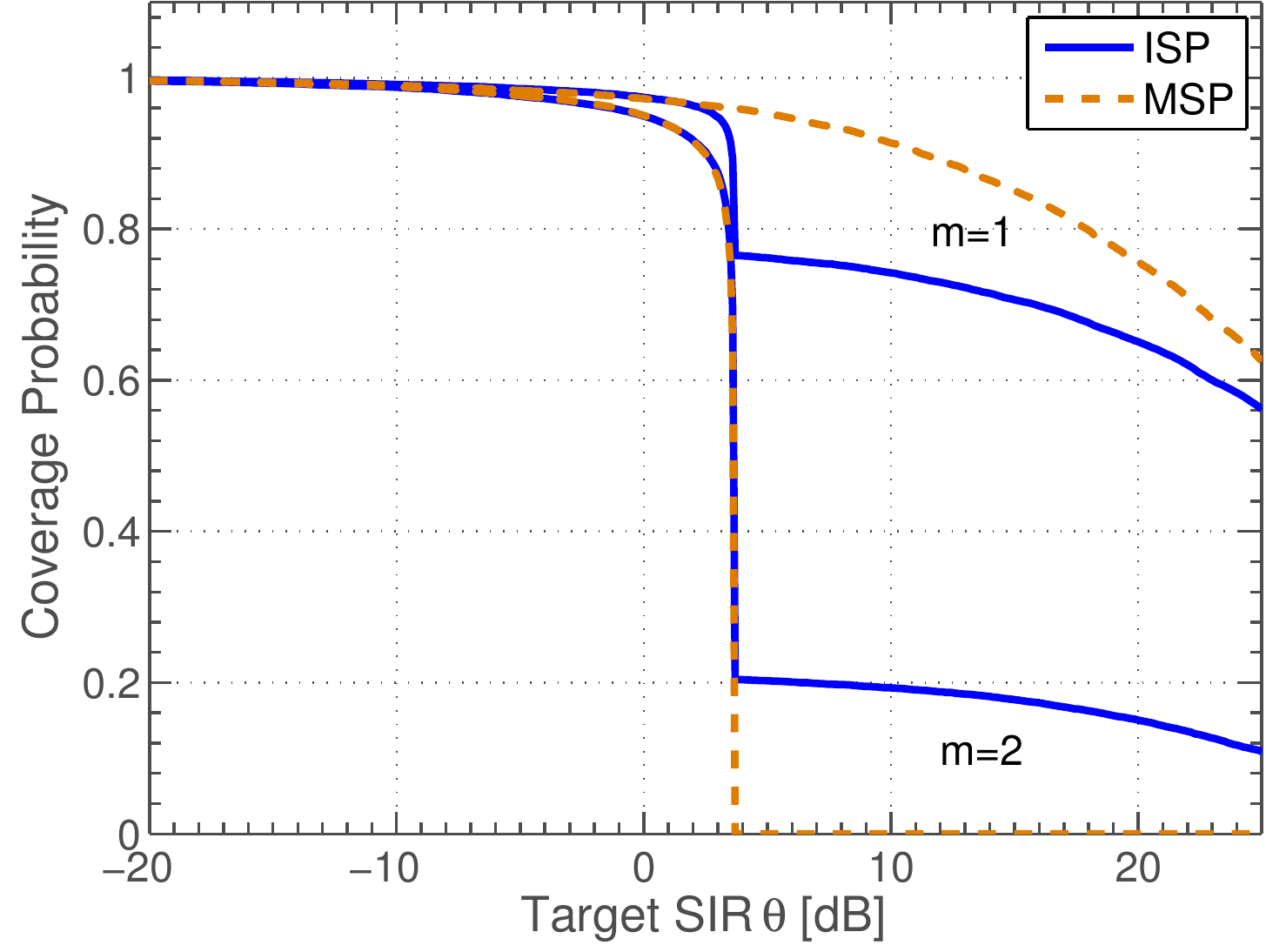,width=.49\textwidth,height=0.9\figwidth}}}} 
	\parbox[c]{.5\textwidth}{%
		\centerline{\subfigure[$\beta=0.5$.]
			{\epsfig{file=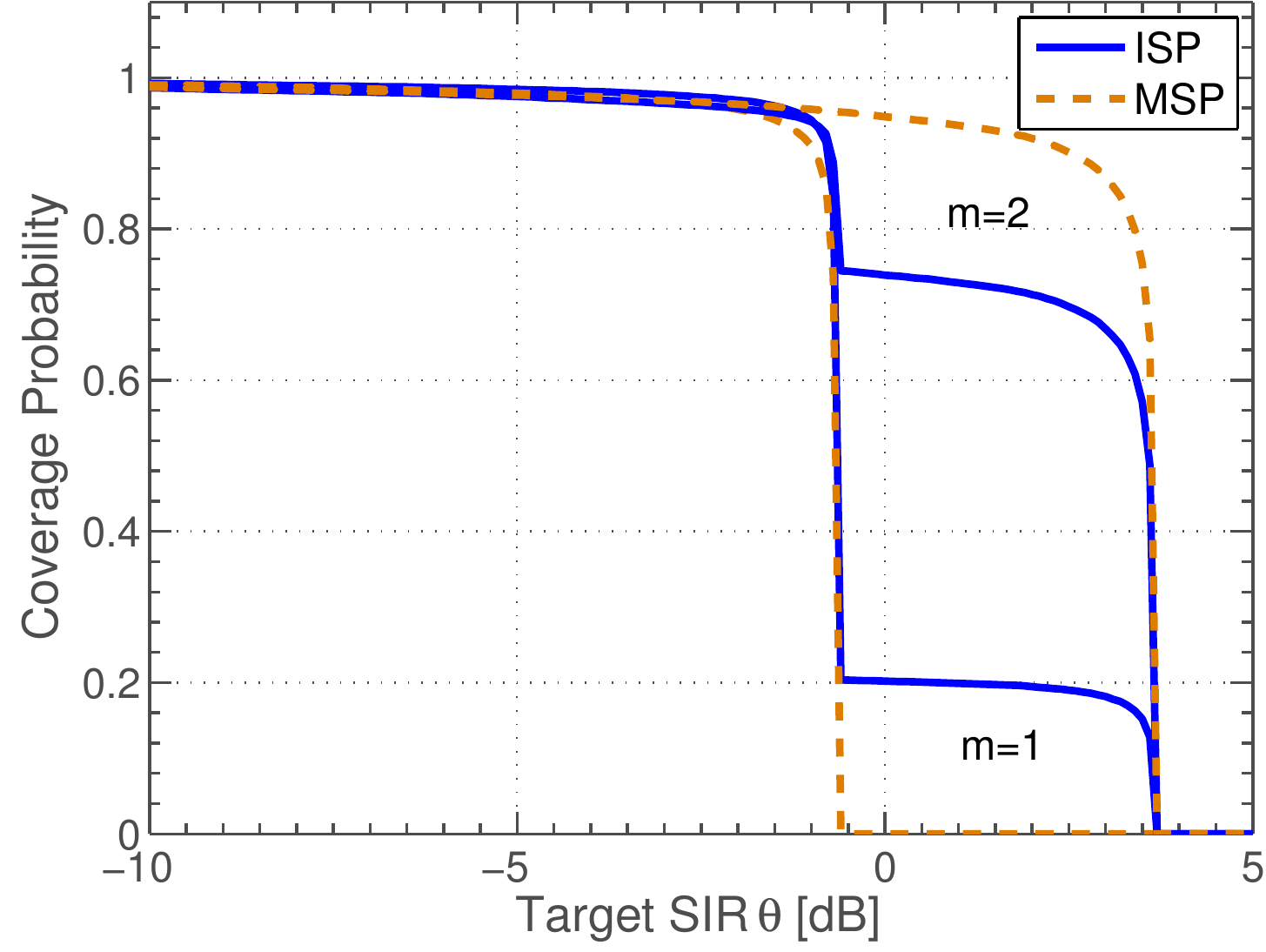,width=.49\textwidth,height=0.9\figwidth}}}} 
	\caption{Downlink coverage probability for MCP model with $N=2$. $\lambda=0.0001$, $R=10$, $\sigma_n^2=0$, $a_1=0.3$, $a_2=0.7$, and $\alpha=4$. Coverage probabilities for instantaneous signal power-based ranking are called ``ISP'' in the legend, and coverage probabilities for distance- (mean signal power-) based ranking are called are called ``MSP''.}
	\label{fig:Coverage_downlink}
\end{figure}
}

\section{Probability of Accuracy for Rayleigh Fading}
In this section, we derive the accuracy probability $\mathcal{A}$ for Rayleigh fading considering PPP, MCP, and TCP for users' spatial location models. {Note that the results provided in this section for Rayleigh fading can also be obtained from the results derived in the next section for Nakagami-$m$ fading by setting $m=1$. However, for $N$-UE NOMA, calculating inner expectation in \eqref{eq:match-expectation} for Nakagami-$m$ fading yields $N-1$ integrals as is shown in {\bf Theorem \ref{Thm4}}. Deriving $\mathcal{A}$ for Rayleigh fading directly from {\bf Definition \ref{Def1}} is easier than deriving from {\bf Theorem \ref{Thm4}} except for some special cases such as $N=2$. Therefore,  we first study the accuracy probability for Rayleigh fading in this section. Since steps of the proofs for Rayleigh and Nakagami-$m$ fading are similar, in the next section, we will only mention the steps or directly provide the final expressions. It is worth mentioning that the results in \textbf{Theorem \ref{Thm2}} and \textbf{Theorem \ref{Thm3}} can be obtained by solving the integrals in \textbf{Theorem \ref{Thm5}} and \textbf{Theorem \ref{Thm6}}, respectively, for $m=1$ and applying binomial expansion.}
Our methodology to derive $\mathcal{A}$ can be described as follows:
\begin{enumerate}
\item Derive the inner expectation in \eqref{eq:match-expectation} by averaging over fading channel powers $\{h_i\}$.
\item Characterize $\mathcal{A}$  for any arbitrary  users'  location model.
\item Derive $\mathcal{A}$ by averaging over the distance distribution of users considering PPP, MCP, and TCP models, respectively.
\end{enumerate}
The first two steps are performed in the following Theorem and the third step is conducted in \textbf{Theorem~2}, \textbf{Theorem~3}, and in subsequent discussions.
\begin{theorem}[Probability of the Accuracy of Distance-based Approximation - $N$ UE NOMA]
\label{Thm1}
For $N$-UE NOMA, the inner expectation in \eqref{eq:match-expectation} can be obtained by
	\begin{IEEEeqnarray}{rCl}
		\mathbb{E}_{\{h_i\}}\left[ \mathbf{1} 
		\left( h_1r_{(1)}^{-\alpha}>h_2r_{(2)}^{-\alpha}>\cdots>h_Nr_{(N)}^{-\alpha} \right) \right] &=&
		\prod_{i=2}^{N} \frac{1}{\sum_{j=1}^i \left(\frac{r_{(j)}}{r_{(i)}}\right)^{\alpha}}.  
		\label{eq:inner-expectation-Rayleigh}
	\end{IEEEeqnarray}
Then, using aforementioned expression, \eqref{eq:inner-expectation-Rayleigh}, and \textbf{Definition~2} we obtain
\begin{IEEEeqnarray}{rCl}
	\mathcal{A}&=&\mathbb{E}\left[ \prod_{i=2}^{N} \frac{1}{\sum_{j=1}^i \left(\frac{r_{(j)}}{r_{(i)}}\right)^{\alpha}} \right] 
	\nonumber \\
	           &=& N! \int_{0}^{\infty} \int_{r_1}^{\infty} \cdots \int_{r_{N-1}}^{\infty}
	           \prod_{i=2}^{N} \frac{1}{\sum_{j=1}^i \left(\frac{r_{j}}{r_{i}}\right)^{\alpha}}
	           f_r(r_1) f_r(r_2) \cdots f_r(r_N) {\rm d}r_N \cdots {\rm d}r_2 {\rm d}r_1.
	\label{eq:final-N_UE-rayleigh}
\end{IEEEeqnarray}  
\end{theorem}
\allowdisplaybreaks{
\begin{IEEEproof}
See {\bf Appendix A.}
\end{IEEEproof}}
According to \textbf{Theorem \ref{Thm1}}, the accuracy probability does not depend on the mean channel power gain $\Omega$. In the following corollary, we provide a simplified expression for the inner expectation in \eqref{eq:match-expectation} considering 2-UE NOMA cluster\footnote{In this paper, we use the term ``$N$-UE NOMA'' to make it explicit that the framework can capture any value of $N$; however, the performance gains of NOMA over OMA (Orthogonal Multiple Access) are generally achievable for small number of user equipment (UE) in a NOMA cluster. Therefore, we are more interested in cases where $N = 2$ and $N = 3$.}.  
\begin{Corollary} \label{Cor1}
Substituting $N=2$ in Theorem \ref{Thm1} and using binomial expansion $(1+x)^{-1}=\sum_{k=0}^{\infty} (-1)^k x^k$, the inner expectation in \eqref{eq:match-expectation} for 2-UE NOMA can be obtained as follows:
	\begin{IEEEeqnarray}{rCl}
		\mathbb{E}_{\{h_i\}}\left[ \mathbf{1} 
		\left( h_1r_{(1)}^{-\alpha}>h_2r_{(2)}^{-\alpha} \right) \right] &=&
		\sum_{k=0}^{\infty} (-1)^k \left(\frac{r_{(1)}}{r_{(2)}}\right)^{\alpha k}. 
		\label{eq:inner-expectation-2_UE-Rayleigh} 
	\end{IEEEeqnarray} 
The accuracy probability can then be derived as follows:
\begin{IEEEeqnarray}{rCl}
	\mathcal{A}=\sum_{k=0}^{\infty} (-1)^k \mathbb{E}\left[  \left(\frac{r_{(1)}}{r_{(2)}}\right)^{\alpha k} \right]
	           =2 \sum_{k=0}^{\infty} (-1)^k \int_{0}^{\infty} \int_{r_1}^{\infty} \left(\frac{r_1}{r_2}\right)^{\alpha k} 
	           f_r(r_1) f_r(r_2) {\rm d}r_2 {\rm d}r_1.
	\label{eq:final-2_UE-rayleigh}
\end{IEEEeqnarray}  
\end{Corollary}
It is worth mentioning that the summation in \eqref{eq:final-2_UE-rayleigh} can be truncated after a few terms since the expression inside the summation is close to zero for large values of $k$. Moreover, unlike the expectation in \eqref{eq:final-N_UE-rayleigh}, the expectation in \eqref{eq:final-2_UE-rayleigh} can be derived in closed-form for PPP, MCP, and TCP models. In the following, we obtain the  accuracy probability in closed-form for each of the models considering two users in a NOMA cluster (i.e., 2-UE NOMA) and then we study the accuracy probability for $N$-UE NOMA. Evidently, for $N$-UE NOMA, there is no closed-form expression available.

\begin{theorem}[Probability of the Accuracy of Distance-based Approximation - 2 UE NOMA and PPP Model]\label{Thm2}
	When each BS serves users that are located in its Voronoi cell, the accuracy probability for 2-UE NOMA is 
	\begin{IEEEeqnarray}{rCl} 
		\mathcal{A}=\sum_{k=0}^{\infty} \frac{(-1)^k}{\alpha k + 2 } \,  _2F_1\left(2,1;\frac{\alpha k}{2}+2;\frac{1}{2} \right).
		\nonumber
	\end{IEEEeqnarray}
\end{theorem}
\begin{IEEEproof} 
See {\bf Appendix~B}.
\end{IEEEproof}
According to \textbf{Theorem \ref{Thm2}}, for $N=2$, the probability that the path-loss-based ranking matches  the instantaneous signal power-based ranking only depends on  path-loss exponent $\alpha$ and does not depend on BS intensity $\lambda$. Similarly, we can generalize Theorem~\ref{Thm2} to the case of $N$-UE NOMA as stated in the following corollary. 
\begin{Corollary}  \label{Cor2}
For $N$-UE NOMA, the accuracy probability $\mathcal{A}$ only depends on path-loss exponent $\alpha$ and does not depend on BS intensity $\lambda$.
\end{Corollary}
\begin{IEEEproof}
See {\bf Appendix~C}.
\end{IEEEproof}

For the MCP model, now we  derive the accuracy probability $\mathcal{A}$ for 2-UE NOMA in closed-form by substituting \eqref{eq:PDF-MCP} in \eqref{eq:final-2_UE-rayleigh}, and for $N$-UE NOMA by substituting  \eqref{eq:PDF-MCP} in \eqref{eq:final-N_UE-rayleigh}. The closed-form expression for the accuracy probability for 2-UE NOMA is provided in the following Theorem.
\begin{theorem}[Probability of the Accuracy of Distance-based Approximation - 2 UE NOMA and MCP Model]\label{Thm3}
When the user point process follows an MCP with parent point process $\Phi$, where $\Phi$ is the BS point process, the accuracy probability of 2-UE NOMA  can be calculated as
	\begin{IEEEeqnarray}{rCl}
		\mathcal{A}=\sum_{k=0}^{\infty} (-1)^k\frac{2}{2+\alpha k}. \nonumber
	\end{IEEEeqnarray}
\end{theorem}
\begin{IEEEproof} The proof follows from substituting \eqref{eq:PDF-MCP} in \eqref{eq:final-2_UE-rayleigh}.
\end{IEEEproof}
Based on \textbf{Theorem \ref{Thm3}}, for 2-UE NOMA, when users are uniformly distributed within distance $R$ from the serving BS, the probability that path-loss-based ranking matches ranking based on the instantaneous signal power only depends on the path-loss exponent $\alpha$ and does not depend on $R$. Next we prove that for $N$-UE NOMA, when users are uniformly distributed within distance $R$ from the serving BS, the accuracy probability does not depend on $R$. By substituting \eqref{eq:PDF-MCP} in \eqref{eq:final-N_UE-rayleigh}, we obtain
	\begin{IEEEeqnarray}{rCl}
		\mathcal{A}=N!\int_0^R\int_{r_1}^R \cdots \int_{r_{N-1}}^R 
		\prod_{i=2}^{N} \frac{1}{\sum_{j=1}^i \left(\frac{r_j}{r_i}\right)^{\alpha}} \frac{2r_1}{R^2} \frac{2r_2}{R^2} \cdots \frac{2r_N}{R^2} {\rm d}r_N \cdots {\rm d}r_2 {\rm d}r_1. 
		\label{eq:step1_Cor3}
	\end{IEEEeqnarray}
Next we simplify the above integral by applying $u_i=\frac{R}{r_i}$ for $i=1,2,...,N$. After changes of variables, the region of integration is: $1<u_1$ and $1<u_i<u_{i-1}$ for $i=2,...,N$. Since the Jacobian matrix $J=\frac{\partial(r_1,\cdots,r_N)}{\partial(u_1,\cdots,u_N)}$ is a diagonal matrix, its determinant is equal to  $\det(J)=\prod_{i=1}^{N} \frac{\partial r_i}{\partial u_i} =\prod_{i=1}^{N} \frac{-R}{u_i^2}$. Therefore, \eqref{eq:step1_Cor3} can be obtained by 
	\begin{IEEEeqnarray}{rCl}
		\mathcal{A}&=& N! 2^N \int_1^{\infty} \int_1^{u_1} \cdots \int_1^{u_{N-1}} u_1^{-3}
		\prod_{i=2}^{N} \frac{u_i^{-3}}{\sum_{j=1}^i \left(\frac{u_i}{u_j}\right)^{\alpha}}
		{\rm d}u_N \cdots {\rm d}u_2 {\rm d}u_1 
		\label{eq:step1_Rayleigh_MCP_N_UE}
	\end{IEEEeqnarray}
which does not depend on $R$. In the following, we further simplify \eqref{eq:step1_Rayleigh_MCP_N_UE} for $N=3$. For other values of $N$, we can also use the same approach. For $N=3$, we have
\begin{IEEEeqnarray}{rCl}
	\mathcal{A}&=& 48 \int_1^{\infty} \int_1^{u_1} \int_1^{u_2} \frac{ u_1^{-3} u_2^{-3} u_3^{-3} }
	{ \left[ 1+\left( \frac{u_2}{u_1}\right)^{\alpha} \right] \left[ 1+\left( \frac{u_3}{u_1}\right)^{\alpha} + \left( \frac{u_3}{u_2}\right)^{\alpha} \right] } {\rm d}u_3 {\rm d}u_2 {\rm d}u_1\nonumber 
	\\
	&\stackrel{\text{(a)}}{=}& 48 \int_1^{\infty} \int_{u_3}^{\infty} \int_{u_2}^{\infty} \frac{ u_1^{-3} u_2^{-3} u_3^{-3} }
	{ \left[ 1+\left( \frac{u_2}{u_1}\right)^{\alpha} \right] \left[ 1+\left( \frac{u_3}{u_1}\right)^{\alpha} + \left( \frac{u_3}{u_2}\right)^{\alpha} \right] } {\rm d}u_1 {\rm d}u_2 {\rm d}u_3 \nonumber 
	\\
	&\stackrel{\text{(b)}}{=}& 48 \int_0^{1} \int_0^1 \int_0^1 \frac{v_1 v_2^3 v_3^5}
	{\left[ 1+v_1^{\alpha} \right] \left[ 1 + v_2^{\alpha} + v_1^{\alpha} v_2^{\alpha} \right]} {\rm d}v_1 {\rm d}v_2 {\rm d}v_3 \nonumber 
	\\
	&=& 8 \int_0^{1} \int_0^1 \frac{v_1 v_2^3}
	{\left[ 1+v_1^{\alpha} \right] \left[ 1 + v_2^{\alpha} + v_1^{\alpha} v_2^{\alpha} \right]} {\rm d}v_1 {\rm d}v_2, 
	\label{eq:accuracy_probability_Rayleigh_MCP_3UE}
    \end{IEEEeqnarray}
where (a) is obtained by changing the orders of the integrals and (b) follows by applying changes of variables $v_3=\frac{1}{u_3}$, $v_2=\frac{u_3}{u_2}$, and $v_1=\frac{u_2}{u_1}$.

\begin{Corollary} \label{Cor3}
When users are uniformly distributed within distance $R$ from the serving BS, for $N$-UE NOMA, the probability that the path-loss-based ranking matches  the instantaneous signal power-based ranking, only depends on the path-loss exponent $\alpha$ and does not depend on $R$.
\end{Corollary}
For the PPP and TCP models, the link distances follow Rayleigh distribution. Therefore, for TCP, we can derive the accuracy probability for 2-UE and $N$-UE simply by replacing $c \lambda \pi$ in \textbf{Theorem \ref{Thm2}} and \textbf{Corollary \ref{Cor2}} with $1/(2 \sigma^2)$. This can be understood by comparing \eqref{eq:PDF-PPP} and \eqref{eq:PDF-TCP}. 
\begin{Corollary} \label{Cor4}
When users are independently and identically distributed with normal distribution with variance $\sigma^2$ around the serving BS, $\mathcal{A}$ for $2$-UE and $N$-UE NOMA can be calculated by \textbf{Theorem \ref{Thm2}} and \textbf{Corollary \ref{Cor2}}, respectively. Therefore, $\mathcal{A}$ is independent of $\sigma^2$. 
\end{Corollary}

\section{Probability of Accuracy for Nakagami-$m$ Fading}
In this section, we derive the probability of accuracy of distance-based approximation considering Nakagami-$m$ fading channels. \textcolor{black}{Using the analytical results, in Section \ref{sec:simulation}, we will show that for more severe fading conditions (i.e., for small values of $m$), the distance-based approximation is less accurate whereas for higher values of $m$, the distance-based approximation is more accurate.} 

Similar to the previous subsection, we first derive the inner expectation in \eqref{eq:match-expectation}. Then the accuracy probability $\mathcal{A}$ is obtained for PPP, MCP, and TCP models as shown in the following.
\begin{theorem}[Probability of the Accuracy of Distance-based Approximation - $N$-UE NOMA] \label{Thm4}
For Nakagami-$m$ fading, with shape parameter $m$, the inner expectation in \eqref{eq:match-expectation} can be derived as follows:
	\begin{IEEEeqnarray}{rCl}
		\IEEEeqnarraymulticol{3}{l}{
		\mathbb{E}_{\{h_i\}}\left[ \mathbf{1} 
		\left( h_1r_{(1)}^{-\alpha}>h_2r_{(2)}^{-\alpha}>\cdots>h_Nr_{(N)}^{-\alpha} \right) \right]} \nonumber \\
	    &=&
		\frac{ \Gamma(Nm) }{ \Gamma(m)^N } 
		\int_1^{\infty} \int_1^{\infty} \cdots \int_1^{\infty} 
		\frac{1}{\left[ 1+\sum_{i=1}^{N-1} \left( \frac{r_{(i)}}{r_{(N)}} \right)^{\alpha} \prod_{k=i}^{N-1}t_k \right]^{Nm}} 
				\prod_{j=1}^{N-1}\left( \frac{r_{(j)}}{r_{(N)}} \right)^{\alpha m}  t_j^{jm-1} {\rm d}t_{N-1} \cdots {\rm d}t_2 {\rm d}t_1. \nonumber 
	\end{IEEEeqnarray}
Then $\mathcal{A}$ can be derived by averaging over the desired link distance distribution using \textbf{Definition~2}.
\end{theorem}
{\allowdisplaybreaks
\begin{IEEEproof} 
See {\bf Appendix~D}.
	\end{IEEEproof}}
Similar to Rayleigh fading, the accuracy probability for Nakagami-$m$ fading does not depend on mean channel power gain. By setting $N=2$ for 2-UE NOMA, we obtain
\begin{IEEEeqnarray}{rCl}
	\mathbb{E}_{\{h_i\}}\left[ \mathbf{1} \left( h_1r_{(1)}^{-\alpha}>h_2r_{(2)}^{-\alpha} \right) \right]
	&=&
	\frac{ \Gamma(2m) }{ \Gamma(m)^2 } \left( \frac{r_{(1)}}{r_{(2)}} \right)^{\alpha m}
	\int_1^{\infty} \frac{t_1^{m-1}  {\rm d}t_1}
	{ \left[1+\left( \frac{r_{(1)}}{r_{(2)}} \right)^{\alpha} t_1 \right]^{2m}  }. 
	\label{eq:inner-expectation_N=2_Nakagami} 
\end{IEEEeqnarray}
For 2-UE NOMA, when $m=1$ (Rayleigh fading), \eqref{eq:inner-expectation_N=2_Nakagami} reverts to \textbf{Corollary \ref{Cor1}}. However, deriving \textbf{Theorem \ref{Thm1}} from \textbf{Theorem \ref{Thm4}} for $N$-UE NOMA, when $m=1$, is not straightforward.

Using \textbf{Definition~2} with \eqref{eq:inner-expectation_N=2_Nakagami} for 2-UE NOMA and with \textbf{Theorem \ref{Thm4}} for $N$-UE NOMA provides the accuracy probability. In the following, similar to the previous section, we derive the accuracy probability for PPP, MCP, and TCP models.
\begin{theorem}[Probability of the Accuracy of Distance-based Approximation - 2 UE NOMA and PPP Model] \label{Thm5}
For 2-UE NOMA and PPP model for users' spatial locations, the accuracy probability for Nakagami-$m$ fading with fading parameter $m$ can be given as follows: 
	\begin{IEEEeqnarray}{rCl}
		\mathcal{A}=\frac{ 4 \Gamma(2m) }{ \Gamma(m) \Gamma(m+1) } 
		\int_0^1 	\frac{u^{1-\alpha m}}{\left( 1 + {u^2}\right)^2} \,_2F_1(2m,m;m+1;-u^{-\alpha})  {\rm d}u. \nonumber  
	\end{IEEEeqnarray}
\end{theorem}
\begin{IEEEproof} 
The accuracy probability can be derived as:
	\begin{IEEEeqnarray}{rCl}
    	\mathcal{A}&=& 	 \frac{ 2 \Gamma(2m) }{ \Gamma(m)^2 } (2 c \lambda \pi)^2 
    	\int_1^{\infty} \int_{0}^{\infty} \int_{r_1}^{\infty} 
    	 \frac{ \left( \frac{r_1}{r_2} \right)^{\alpha m}  t_1^{m-1} }
    	{ \left[1+\left( \frac{r_1}{r_2} \right)^{\alpha} t_1 \right]^{2m}  } 
    	r_1 r_2 e^{-c \lambda \pi \left( r_1^2 + r_2^2\right)}  {\rm d}r_2 {\rm d}r_1 {\rm d}t_1 
    	\nonumber \\
    	&\stackrel{\text{(a)}}{=}& \frac{ 2 \Gamma(2m) }{ \Gamma(m)^2 } (2 c \lambda \pi)^2 
    	\int_1^{\infty} \int_{0}^{1} 
    	\frac{  u^{\alpha m-3}  t_1^{m-1} }
    	{ \left[1+u^{\alpha} t_1 \right]^{2m}  } 
        \int_{0}^{\infty} v^3	e^{-c \lambda \pi \left( 1 + \frac{1}{u^2}\right)v^2}   {\rm d}v {\rm d}u {\rm d}t_1 
    	\nonumber \\
    	&\stackrel{\text{(b)}}{=}& \frac{ 4 \Gamma(2m) }{ \Gamma(m)^2 } 
    	\int_1^{\infty} \int_{0}^{1} 
    	\frac{  u^{\alpha m+1}  t_1^{m-1} }
    	{ \left[1+u^{\alpha} t_1 \right]^{2m}  } 
    	\frac{1}{\left( 1 + {u^2}\right)^2} {\rm d}u {\rm d}t_1 
    	\nonumber \\
    	&\stackrel{\text{(c)}}{=}& \frac{ 4 \Gamma(2m) }{ \Gamma(m)^2 } 
    	\int_0^1 	\frac{u^{1-\alpha m}}{\left( 1 + {u^2}\right)^2} \int_0^1 
    	\frac{ z^{m-1} } { \left[1+u^{-\alpha} z \right]^{2m}  } {\rm d}z {\rm d}u,  
    	\nonumber
	\end{IEEEeqnarray}
where (a) is obtained by changes of variables $\frac{r_1}{r_2}=u$ and $r_1=v$. (b) follows by applying $c \lambda \pi \left( 1 + \frac{1}{u^2}\right)v^2=x$. (c) is obtained by $t_1^{-1}=z$. Finally, \textbf{Theorem \ref{Thm5}} can be derived by using the integral representation of Gaussian hypergeometric function.
\end{IEEEproof}
According to \textbf{Theorem \ref{Thm5}}, the accuracy probability in Nakagami-$m$ fading for 2-UE NOMA does not depend on the BS intensity $\lambda$. In the following, we prove that, for $N$-UE NOMA with Nakagami-$m$ fading, the accuracy probability is independent of $\lambda$.
\begin{Corollary} \label{Cor5}
For $N$-UE NOMA and PPP model, the accuracy probability for Nakagami-$m$ fading with  parameter $m$ is independent of the BS intensity $\lambda$. 
\end{Corollary} 
\begin{IEEEproof}
	See {\bf Appendix~E}.
\end{IEEEproof}
Now we derive the accuracy probability $\mathcal{A}$ for MCP model by averaging \eqref{eq:inner-expectation_N=2_Nakagami} (for 2-UE NOMA) and \textbf{Theorem \ref{Thm4}} (for $N$-UE NOMA) with respect to $\{r_{(i)}\}$, where the joint PDF $f_{ r_{(1)},r_{(2)},\cdots,r_{(N)} }(x_1,x_2,\cdots,x_N)$ can be obtained by substituting \eqref{eq:PDF-MCP} in \textbf{Definition~2}. 

\begin{theorem}[Probability of the Accuracy of Distance-based Approximation - 2 UE NOMA and MCP Model] \label{Thm6}
	When user point process follows an MCP with parent point process $\Phi$, where $\Phi$ is the BS point process, the accuracy probability of 2-UE NOMA in Nakagami-$m$ fading with parameter $m$ is as follows:
	\begin{IEEEeqnarray}{rCl}
		\mathcal{A}=\frac{ 2 \Gamma(2m) }{ \Gamma(m) \Gamma(m+1) } 
		\int_0^1 	{u^{1-\alpha m}} \,_2F_1(2m,m;m+1;-u^{-\alpha})  {\rm d}u. \nonumber  
	\end{IEEEeqnarray}
\end{theorem}
\begin{IEEEproof} 
	See {\bf Appendix~F}.
\end{IEEEproof}
According to \textbf{Theorem \ref{Thm6}}, for 2-UE NOMA, $\mathcal{A}$ does not depend on $R$ in MCP model. For $N$-UE NOMA, we can also prove that the accuracy probability is independent of $R$ as stated in the following corollary. 
\begin{Corollary} \label{Cor6}
For Nakagami-$m$ fading, when users are uniformly distributed within distance $R$ from the serving BS, for $N$-UE NOMA, the probability that path-loss-based ranking matches the instantaneous signal power-based ranking  does not depend on $R$.
\end{Corollary}
\begin{IEEEproof}
	The proof follows from \eqref{eq:final-N_UE-Nakagami}, where to solve the expectation, we can use the same changes of variables as we used to simplify \eqref{eq:step1_Cor3}:
	    \begin{IEEEeqnarray}{rCl}
		\IEEEeqnarraymulticol{3}{l}{	\mathbb{E} \left[ 
			\frac{ \prod_{j=1}^{N-1}\left( \frac{r_{(j)}}{r_{(N)}} \right)^{\alpha m} }
			{\left[ 1+\sum_{i=1}^{N-1} \left( \frac{r_{(i)}}{r_{(N)}} \right)^{\alpha} \prod_{k=i}^{N-1}t_k \right]^{Nm}} \right] =} 
		\nonumber \\
		&&\> N!\,2^{N} \int_{1}^{\infty} \int_{1}^{u_1} \cdots \int_{1}^{u_{N-1}} 
		\frac{ \prod_{j=1}^{N-1}\left( \frac{u_N}{u_j} \right)^{\alpha m+3} }
		{\left[ 1+\sum_{i=1}^{N-1} \left( \frac{u_N}{u_i} \right)^{\alpha} \prod_{k=i}^{N-1}t_k \right]^{Nm}} u_N^{-3N}
		{\rm d}{u_N} \cdots {\rm d}{u_2} {\rm d}{u_1}. \nonumber 
	\end{IEEEeqnarray}
    The above equation can be further simplified similar to \eqref{eq:accuracy_probability_Rayleigh_MCP_3UE}.
\end{IEEEproof}
Finally, for the TCP model, we can derive the accuracy probability by replacing $c \lambda \pi$ in the final expressions of the accuracy probability of PPP model with $1/( 2 \sigma^2 )$ . Since $c \lambda \pi$ cancels out in the final expressions, \textbf{Theorem \ref{Thm5}} and \textbf{Corollary \ref{Cor5}} are also applicable for TCP. Moreover, according to \textbf{Corollary \ref{Cor5}}, we can conclude that the accuracy probability does not depend on $\sigma^2$.

\section{User Pairing and Probability of Accuracy}
In the previous sections, from the set of users that are associated to the same BS, $N$ users were randomly selected to form a NOMA cluster. However, in practice, NOMA users are chosen such that NOMA gain can be achieved over OMA. For instance, to form a 2-UE NOMA cluster, out of $M$ users associated to the typical BS, usually the nearest and the farthest users are selected. In the following, we study the accuracy probability with user pairing. 

To form a NOMA cluster, we have selected $N$ users from $M$ users that are associated to the typical BS. We denote rank of the selected users by the set $s=\{ s_{(i)} \}$, where $i=1,2,\cdots,N$, $s_{(i)}\in\left\{1,2,...,M\right\}$, and $1 \le s_{(1)} < s_{(2)} \cdots < s_{(N-1)} < s_{(N)} \le M $. From \textbf{Definition~2} and \textbf{Theorem \ref{Thm1}}, for Rayleigh fading, we obtain 
\begin{IEEEeqnarray}{rCl}
	\mathcal{A}&=&\mathbb{E}\left[ \prod_{i=2}^{N} \frac{1}{\sum_{j=1}^i \left(\frac{r_{s_{(j)}}}{r_{s_{(i)}}}\right)^{\alpha}} \right] 
	\nonumber \\
	&=& M! \int_{0}^{\infty} \int_{r_1}^{\infty} \cdots \int_{r_{M-1}}^{\infty}
	\prod_{i=2}^{N} \frac{1}{\sum_{j=1}^i \left(\frac{r_{s_{(j)}}}{r_{s_{(i)}}}\right)^{\alpha}}
	f_r(r_1) f_r(r_2) \cdots f_r(r_M) {\rm d}r_M \cdots {\rm d}r_2 {\rm d}r_1.
	\label{eq:final-N_UE-rayleigh-UE_Pairing}
\end{IEEEeqnarray}
For $N=2$, when we select the nearest and the farthest users, i.e, $s_{(1)}=1$ and $s_{(2)}=M$, \eqref{eq:final-N_UE-rayleigh-UE_Pairing} can be simplified as in the following:
\begin{IEEEeqnarray}{rCl}
	\mathcal{A}
	&=& M! \int_{0}^{\infty} \int_{r_1}^{\infty} \cdots \int_{r_{M-1}}^{\infty}
	\frac{1}{1+\left(\frac{r_1}{r_M}\right)^{\alpha}}
	f_r(r_1) f_r(r_2) \cdots f_r(r_M) {\rm d}r_M \cdots {\rm d}r_2 {\rm d}r_1 \nonumber \\
	&\stackrel{\text{(a)}}{=}& \frac{M!}{(M-2)!} \int_{0}^{\infty} \int_{r_1}^{\infty} 
	\frac{1}{1+\left(\frac{r_1}{r_M}\right)^{\alpha}} \left[F_r(r_M)-F_r(r_1)\right]^{M-2}
	f_r(r_1) f_r(r_M) {\rm d}r_M {\rm d}r_1,
	\label{eq:final-N_UE-rayleigh-2UE_Pairing}
\end{IEEEeqnarray}
\textcolor{black}{where (a) is obtained using the technique  in \cite{ahsanullah2013introduction} to derive Equation 2.12, i.e., for i.i.d. random variables $r_2,r_3,\cdots,r_{M-1}$, $\left[F_r(r_M)-F_r(r_1)\right]^{M-2}$ is the probability that they are in the interval $[r_1,r_M]$. Sorting these random variables in an ascending order based on their realizations gives $(M-2)!$ different permutations out of which only one satisfies the condition $r_2<r_3<\cdots<r_{M-1}$}. We can similarly simplify \eqref{eq:final-N_UE-rayleigh-UE_Pairing} for other values of $N$ and different selection of NOMA users.
Note that the same result can also be obtained by averaging the result in \textbf{Theorem \ref{Thm1}} with respect to the joint PDF of  $r_{s_{(1)}}, r_{s_{(2)}}, \cdots, r_{s_{(N)}}$, which is also provided in \cite{ahsanullah2013introduction}. 
\begin{Corollary}\label{Cor_2UEpairing}
 For 2-UE NOMA, the accuracy probability, when we select the nearest and the farthest users, i.e., $s_{(1)}=1$ and $s_{(2)}=M$, is an increasing function of $M$ irrespective of the fading channel and users' spatial distributions.
\end{Corollary}
\begin{IEEEproof}
See {\bf Appendix G.}	
\end{IEEEproof}
For Nakagami-$m$ fading, using \textbf{Definition~2} and \textbf{Theorem \ref{Thm4}} gives $\mathcal{A}=$
\begin{IEEEeqnarray}{rCl}
	\frac{ \Gamma(Nm) }{ \Gamma(m)^N } 
	\int_1^{\infty} \int_1^{\infty} \cdots \int_1^{\infty} 
	\mathbb{E} \left[ 
	\frac{ \prod_{j=1}^{N-1}\left( \frac{r_{s_{(j)}}}{r_{s_{(N)}}} \right)^{\alpha m} }
	{\left[ 1+\sum_{i=1}^{N-1} \left( \frac{r_{s_{(i)}}}{r_{s_{(N)}}} \right)^{\alpha} \prod_{k=i}^{N-1}t_k \right]^{Nm}} \right]
	\prod_{j=1}^{N-1}  t_j^{jm-1} {\rm d}t_{N-1} \cdots {\rm d}t_2 {\rm d}t_1, \nonumber \\
	\label{eq:eq:final-N_UE-Nakagami_UE_Pairing}
\end{IEEEeqnarray}
where 
\begin{IEEEeqnarray}{rCl}
	\IEEEeqnarraymulticol{3}{l}{
    \mathbb{E} \left[ \frac{ \prod_{j=1}^{N-1}\left( \frac{r_{s_{(j)}}}{r_{s_{(N)}}} \right)^{\alpha m} }
    {\left[ 1+\sum_{i=1}^{N-1} \left( \frac{r_{s_{(i)}}}{r_{s_{(N)}}} \right)^{\alpha} \prod_{k=i}^{N-1}t_k \right]^{Nm}} \right] =}
    \nonumber \\
 M!\int_0^{\infty} \int_{r_1}^{\infty} \cdots \int_{r_{M-1}}^{\infty} 
	\frac{ \prod_{j=1}^{N-1}\left( \frac{r_{s_{(j)}}}{r_{s_{(N)}}} \right)^{\alpha m} }
	{\left[ 1+\sum_{i=1}^{N-1} \left( \frac{r_{s_{(i)}}}{r_{s_{(N)}}} \right)^{\alpha} \prod_{k=i}^{N-1}t_k \right]^{Nm}} 
	f_r(r_1)f_r(r_2) \cdots f_r(r_M) {\rm d}r_{M} \cdots {\rm d}r_2 {\rm d}r_1. \nonumber \\
	\label{eq:eq:expectation-N_UE-Nakagami_UE_Pairing}
\end{IEEEeqnarray}
Now using the above equations, we can study $\mathcal{A}$ for different users' location models.
\begin{Corollary}
When each BS serves users  in its Voronoi cell (PPP model), for any selection of users for the NOMA cluster, the accuracy probability is independent of BS intensity $\lambda$.
\end{Corollary}
\begin{IEEEproof}
The proof can be obtained by using the same approach as in the proof of \textbf{Corollary \ref{Cor2}} for \eqref{eq:final-N_UE-rayleigh-UE_Pairing} (Rayleigh fading) and \eqref{eq:eq:expectation-N_UE-Nakagami_UE_Pairing} (for Nakagami-$m$ fading). 
\end{IEEEproof}
\begin{Corollary}When users are uniformly distributed within distance $R$ from the serving BS (MCP model) and for any user selection scheme, the accuracy probability is independent of $R$.
\end{Corollary}
\begin{IEEEproof}
	The proof can be obtained by using the same approach as in the proof of Corollary \ref{Cor3} for \eqref{eq:final-N_UE-rayleigh-UE_Pairing} (Rayleigh fading) and \eqref{eq:eq:expectation-N_UE-Nakagami_UE_Pairing} (for Nakagami fading). 
\end{IEEEproof}
\begin{Corollary}
	When users are independently and identically scattered with normal distribution with variance $\sigma^2$ around the serving BS (TCP model), for any selection of users for the NOMA cluster, the accuracy probability is independent of $\sigma^2$.
\end{Corollary}

\section{Numerical and Simulation Results} \label{sec:simulation}
This section demonstrates the efficacy of the derived expressions by comparing them to Monte-Carlo simulations.
In Table \ref{my-label1}, we summarize the expressions defining the accuracy of the distance-based approximation in NOMA assuming different spatial  models and fading models.  We use Gaussian quadrature method to approximate and solve four or higher dimensional integrals. In the following, we briefly review the Gaussian quadrature method, describe simulation parameters, and then present our results which demonstrate the impact of path-loss exponent, fading parameter $m$ (in Nakagami-$m$ fading), and user pairing on the accuracy probability. 
\begin{table*}[!ht]
	\centering
	\caption{Accuracy Probability $\mathcal{A}$ for Random User Selection}
	\label{my-label1}
	\begin{tabular}{|p{1.5cm}|p{2.5cm}|p{0.5cm}|p{10cm}|}
		\hline
		{\bf Fading} & { \bf Netwrok Model} & {\bf  $N$}  & {\bf Accuracy Probability ($\mathcal{A}$)}
		\vspace{3mm}
		\\ 
		\hline
		Rayleigh  & PPP/TCP & 2 & 
        $\sum_{k=0}^{\infty} \frac{(-1)^k}{ 2 + \alpha k } \,  _2F_1\left(2,1;\frac{\alpha k}{2}+2;\frac{1}{2} \right)$
		\\
		\hline
		Rayleigh  & PPP/TCP & 3 &
		$48\int_0^1 \int_0^1 \frac{u_1 u_2^3}
		{ \left( 1+u_1^{\alpha} \right)\left( 1+u_2^{\alpha}+u_1^{\alpha}u_2^{\alpha} \right)\left( 1+u_2^2+u_1^2u_2^2 \right)^3 } 
		{\rm d}u_2 {\rm d}u_1$
		\\ 
		\hline
		Raylegih  & MCP     & 2 &
		$\sum_{k=0}^{\infty} (-1)^k \frac{2}{ 2 + \alpha k }$ 
		\\
		\hline
		Rayleigh  & MCP      & 3 &
		$ 8\int_0^1 \int_0^1 \frac{u_1 u_2^3}
		{ \left( 1+u_1^{\alpha} \right)\left( 1+u_2^{\alpha}+u_1^{\alpha}u_2^{\alpha} \right) } 
		{\rm d}u_2 {\rm d}u_1$
		\\
		\hline
		Nakagami & PPP/TCP & 2 &  
		$\frac{ 4 \Gamma(2m) }{ \Gamma(m) \Gamma(m+1) } 
		\int_0^1 	\frac{u^{1-\alpha m}}{\left( 1 + {u^2}\right)^2} \,_2F_1(2m,m;m+1;-u^{-\alpha})  {\rm d}u$
		\\
		\hline
		Nakagami & PPP/TCP & 3 & 
		$\frac{ 48 \Gamma(3m) }{ \Gamma(m)^3 }  
		\int_0^1 \int_0^1 \int_0^1 \int_0^1 
		\frac{u_1^{1+\alpha m} u_2^{3+2\alpha m} z_1^{2m-1} z_2^{m-1}}
		{\left( u_1^{\alpha} u_2^{\alpha} + u_2^{\alpha} z_1 + z_1 z_2 \right)^{3m} \left( 1 + u_2^2 + u_1^2 u_2^2\right)^3} {\rm d}u_2 {\rm d}u_1 {\rm d}z_2 {\rm d}z_1$
		\\
		\hline
		Nakagami & MCP     & 2 &  
		$\frac{ 2 \Gamma(2m) }{ \Gamma(m) \Gamma(m+1) } 
		\int_0^1 	u^{1-\alpha m} \,_2F_1(2m,m;m+1;-u^{-\alpha})  {\rm d}u$
		\\
		\hline
		Nakagami & MCP     & 3 & 
		$\frac{ 8 \Gamma(3m) }{ \Gamma(m)^3 }  
		\int_0^1 \int_0^1 \int_0^1 \int_0^1 
		\frac{u_1^{1+\alpha m} u_2^{3+2\alpha m} z_1^{2m-1} z_2^{m-1}}
		{\left( u_1^{\alpha} u_2^{\alpha} + u_2^{\alpha} z_1 + z_1 z_2 \right)^{3m} } {\rm d}u_2 {\rm d}u_1 {\rm d}z_2 {\rm d}z_1$ 
		\\
		\hline
	\end{tabular}
\end{table*}

\subsection{Approximation of Multi-Dimensional Integrals}
A quadrature rule provides an approximation of the definite integral of a function, usually stated as a weighted sum of function values at specified points within the domain of integration.  
\begin{Definition}[Gaussian Quadrature]
When domain of integration is $[0,1]$\footnote{Note that domains of integrals in Table \ref{my-label1} are all $[0,1]$.}, an n-point Gaussian quadrature rule states
\begin{IEEEeqnarray}{rCl}
	\int_0^1 f(x) {\rm d}x \approx \sum_{i=1}^n w_i f(x_i), \nonumber
\end{IEEEeqnarray}
where the weights $w_i$ and nodes $x_i$ are obtained such that the approximation is exact for a set of $2n$ different functions \cite{stroud1966gaussian}.
\end{Definition}
To evaluate the four dimensional integrals in Table \ref{my-label1}, we  use the following approximation:
\begin{IEEEeqnarray}{rCl}
	\int_0^1 \int_0^1 \int_0^1 \int_0^1 f(x_1,x_2,x_3,x_4) {\rm d}x_1 {\rm d}x_2 {\rm d}x_3 {\rm d}x_4 \approx 
	\sum_{i_1=1}^n \sum_{i_2=1}^n \sum_{i_3=1}^n \sum_{i_4=1}^n w_{i_1} w_{i_2} w_{i_3} w_{i_4} f(x_{i_1},x_{i_2},x_{i_3},x_{i_4}), \nonumber
\end{IEEEeqnarray} 
where 30-point ($n=30$) Gaussian quadrature rule is employed. The values of the weights $w_i$ and nodes $x_i$ are provided in \cite{ma1996generalized}[Table 3].

\subsection{Simulation Parameters}
We consider $\lambda=0.0005$, $R=20$, $\sigma^2$ = 25, and $\Omega$   = 1. 
Note that when the numerical results match the simulation results, the numerical results are presented. As we have mentioned in Section II, \eqref{eq:PDF-PPP} is an approximation for the PDF of the desired link distance of the typical Voronoi cell. Therefore, for the PPP model, we plot both numerical and simulation results. Moreover, as we have mentioned in the previous subsection, four or higher dimensional integrals are approximated using generalized Gaussian quadrature method. Hence, for $N=3$ in Nakagami-$m$ fading, simulation and analytical results for all PPP, TCP, and MCP models are also provided.

\subsection{Results and Discussions}

\subsubsection{Impact of Path-Loss Exponent}
\begin{figure}
	\centering
	\includegraphics[width=1.25\figwidth]{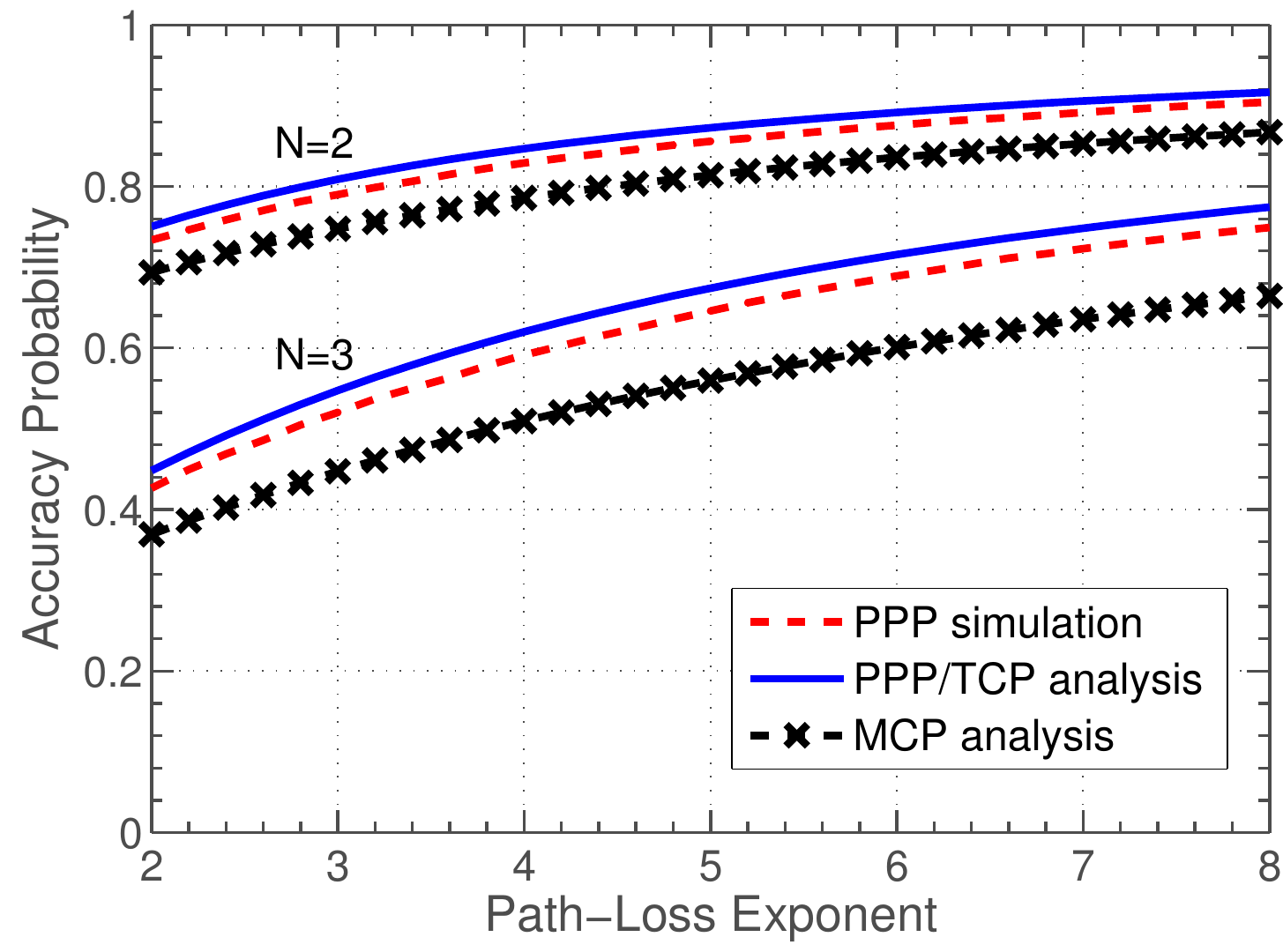}
	\caption{Accuracy probability as a function of path-loss exponent for Rayleigh fading.}
	\label{fig:AvsPLE_Rayleigh}	
\end{figure}

\begin{figure}
	\parbox[c]{.5\textwidth}{%
		\centerline{\subfigure[PPP and TCP.]
			{\epsfig{file=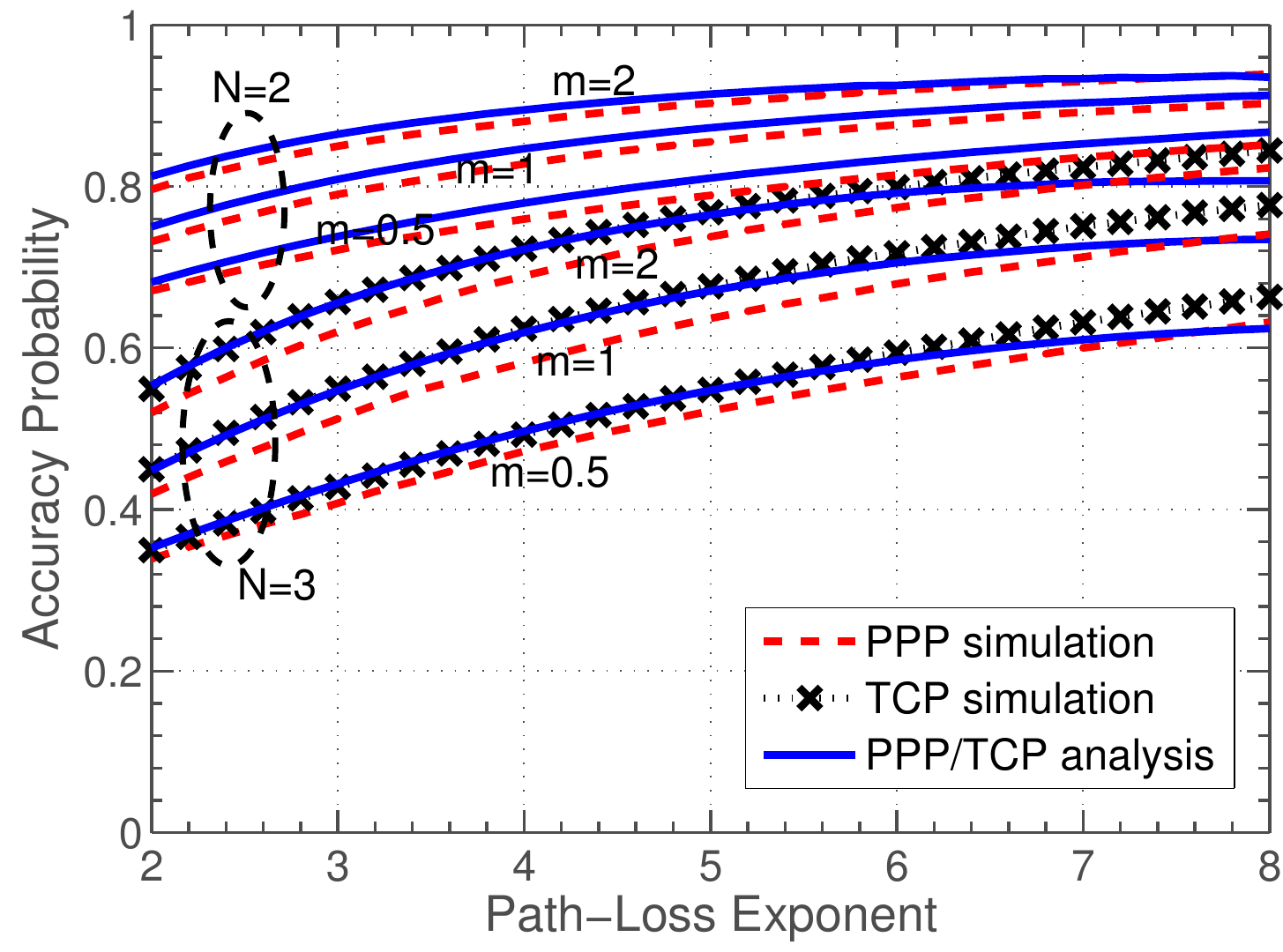,width=.49\textwidth,height=0.9\figwidth}}}} 
	\parbox[c]{.5\textwidth}{%
		\centerline{\subfigure[MCP.]
			{\epsfig{file=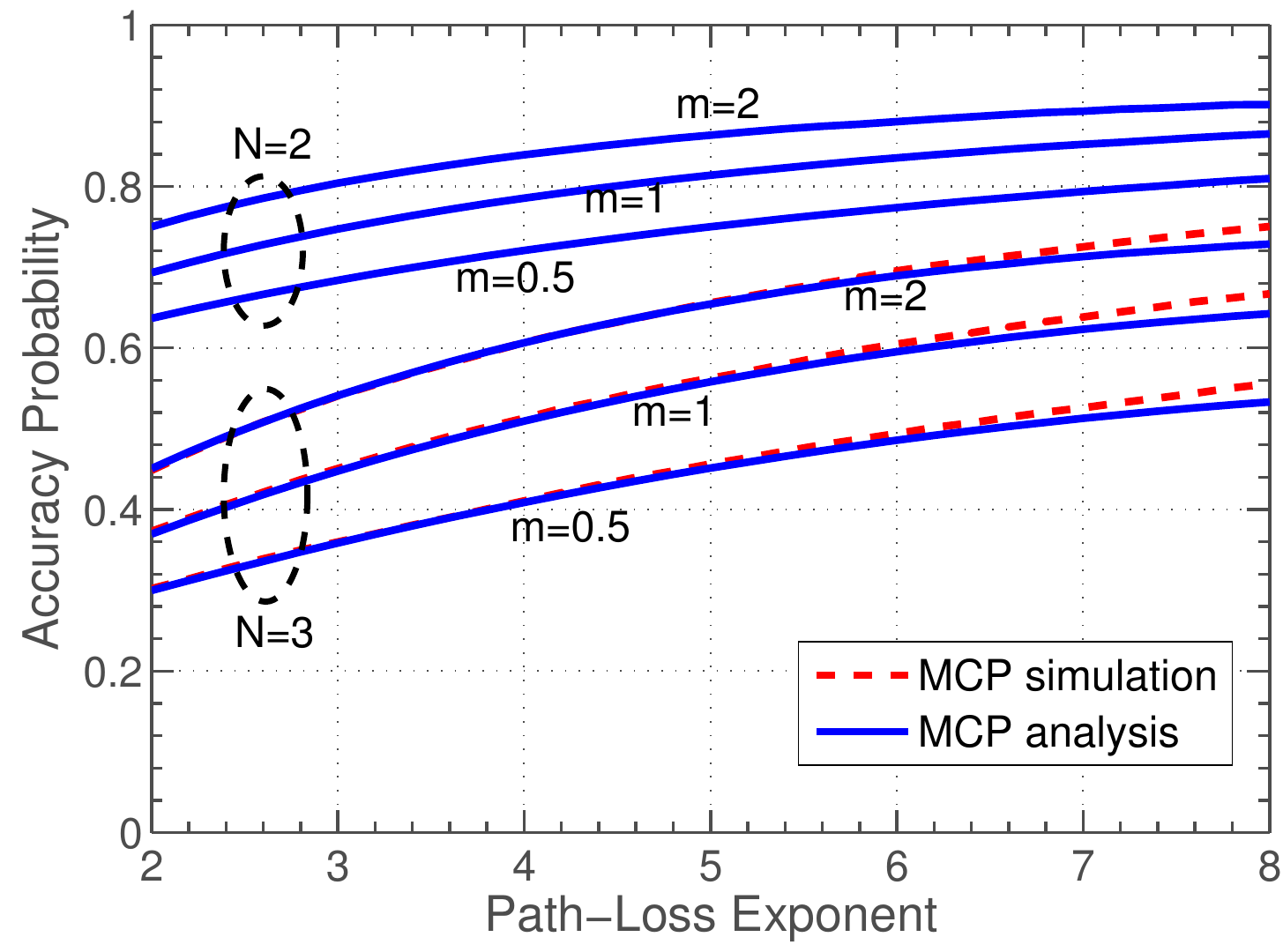,width=.49\textwidth,height=0.9\figwidth}}}} 
	\caption{Accuracy probability as a function of path-loss exponent for Nakagami-$m$ fading with $m=0.5,1,2$.}
	\label{fig:AvsPLE_Nakagami}
\end{figure}
In \figref{fig:AvsPLE_Rayleigh} and \figref{fig:AvsPLE_Nakagami}, the accuracy probability for Rayleigh and Nakagami-$m$ fading is illustrated as a function of path-loss exponent $\alpha$. The analytical results are provided in Table \ref{my-label1} for two and three users. According to \figref{fig:AvsPLE_Rayleigh}, for Rayleigh fading with $\alpha=4$, ranking users based on their distances for 2-UE NOMA is accurate with probability 0.84 for PPP and TCP. For MCP, ranking users based on their distances is valid with probability 0.79. Therefore, for $N=2$, ordering users based on their distances instead of instantaneous signal powers seems reasonable. However, for $N=3$, accuracy probability decreases significantly. When $\alpha=4$, the accuracy probability is about 0.61 for TCP and PPP, and is 0.51 for MCP. 

In \figref{fig:AvsPLE_Nakagami}, for Nakagami-$m$ fading, the accuracy probability is illustrated for different values of $m$. For $N=3$, we use the Gaussian quadrature method to numerically evaluate the four dimensional integrals in Table \ref{my-label1}. Note that, for $N=3$, the difference between simulation results and analysis for TCP and MCP in \figref{fig:AvsPLE_Nakagami}(a) and  \figref{fig:AvsPLE_Nakagami}(b) is due to the Gaussian quadrature method. In summary, we can observe that distance-based ranking  yields more accurate coverage probability results for higher values of $\alpha$, $m$, and less number of users in a NOMA cluster.

\subsubsection{Impact of Fading Parameter $m$}
\begin{figure}
	\centering
	\includegraphics[width=1.25\figwidth]{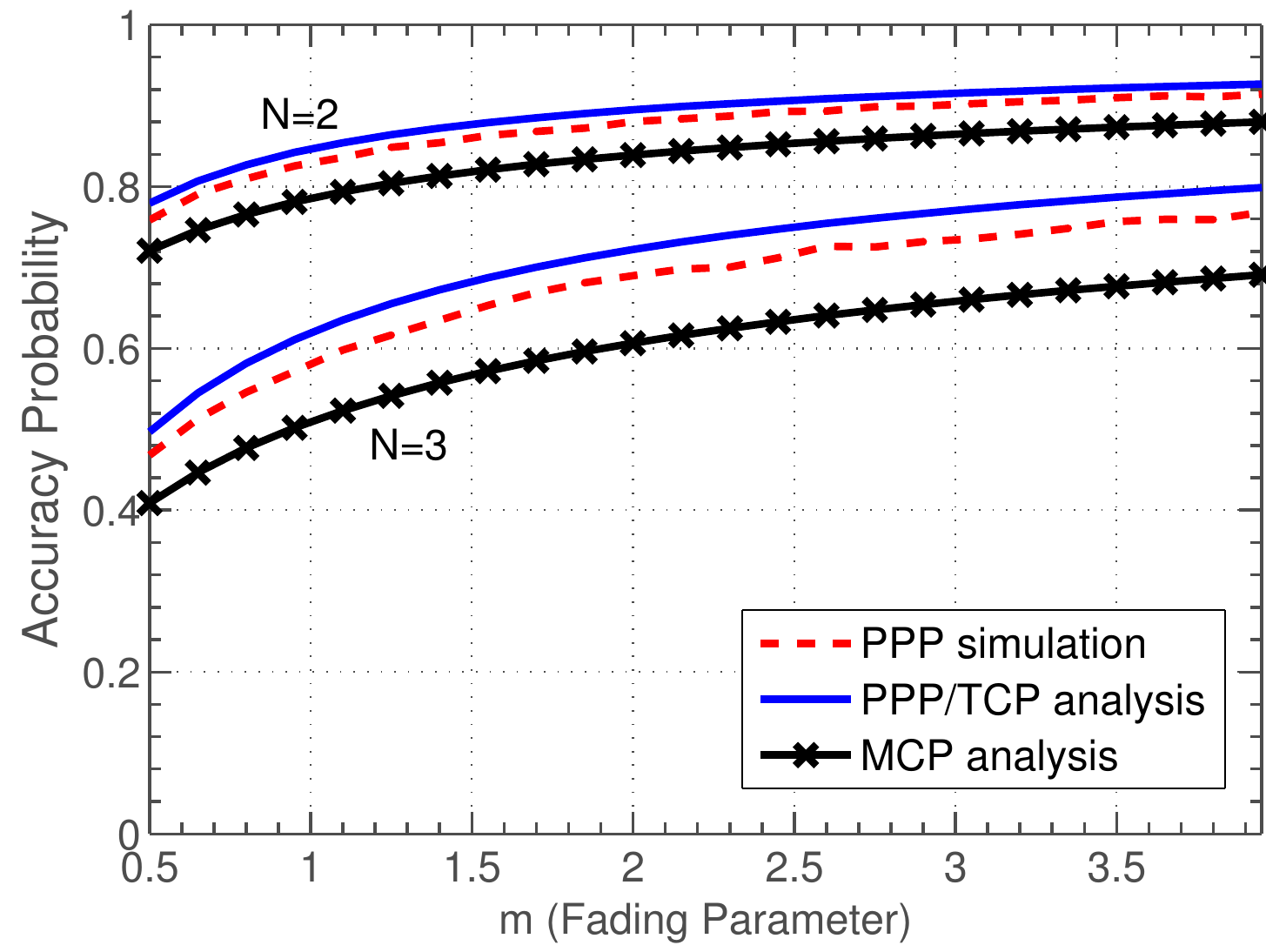}
	\caption{Accuracy probability as a function of fading parameter $m$ for $\alpha=4$.}
	\label{fig:Avsm}	
\end{figure}
When $\alpha=4$, for Nakagami-$m$ fading, in \figref{fig:Avsm}, the accuracy probability is shown as a function of $m$. As we can see, the accuracy probability is an increasing function of $m$. Therefore, in scenarios  with better fading conditions, the distance-based approximation is reasonable. This result is also intuitive because when fading conditions improve, the impact of fading on the channel power is not  significant and the distance-based path-loss is dominant. As such, the distance-based approximation is reasonable. 

\subsubsection{Impact of Distance-Based User Selection}
\begin{figure}
	\parbox[c]{.5\textwidth}{%
		\centerline{\subfigure[PPP and 
TCP.]{\epsfig{file=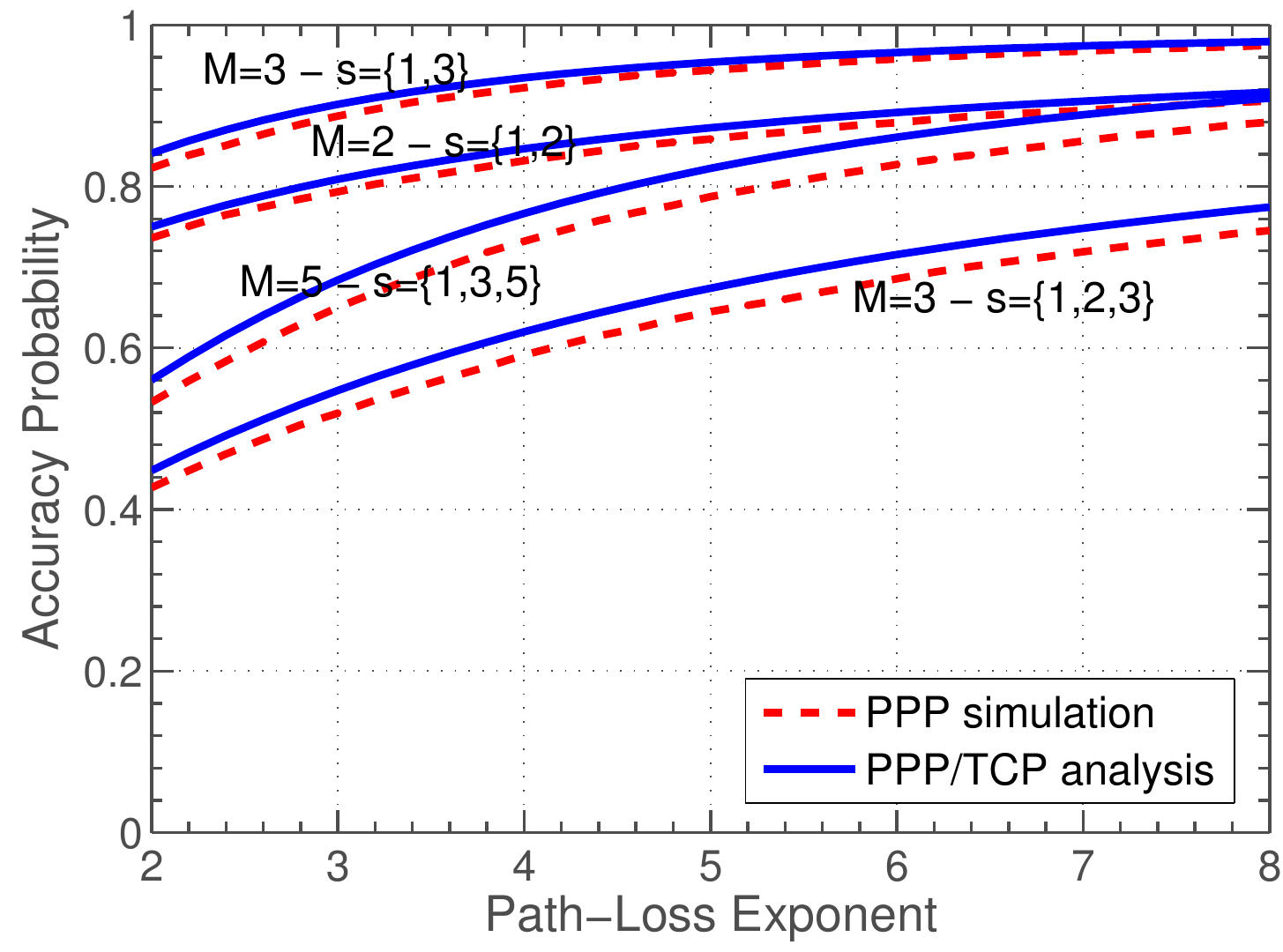,width=.49\textwidth,height=0.9\figwidth}}}} 
	\parbox[c]{.5\textwidth}{%
		\centerline{\subfigure[MCP.]
{\epsfig{file=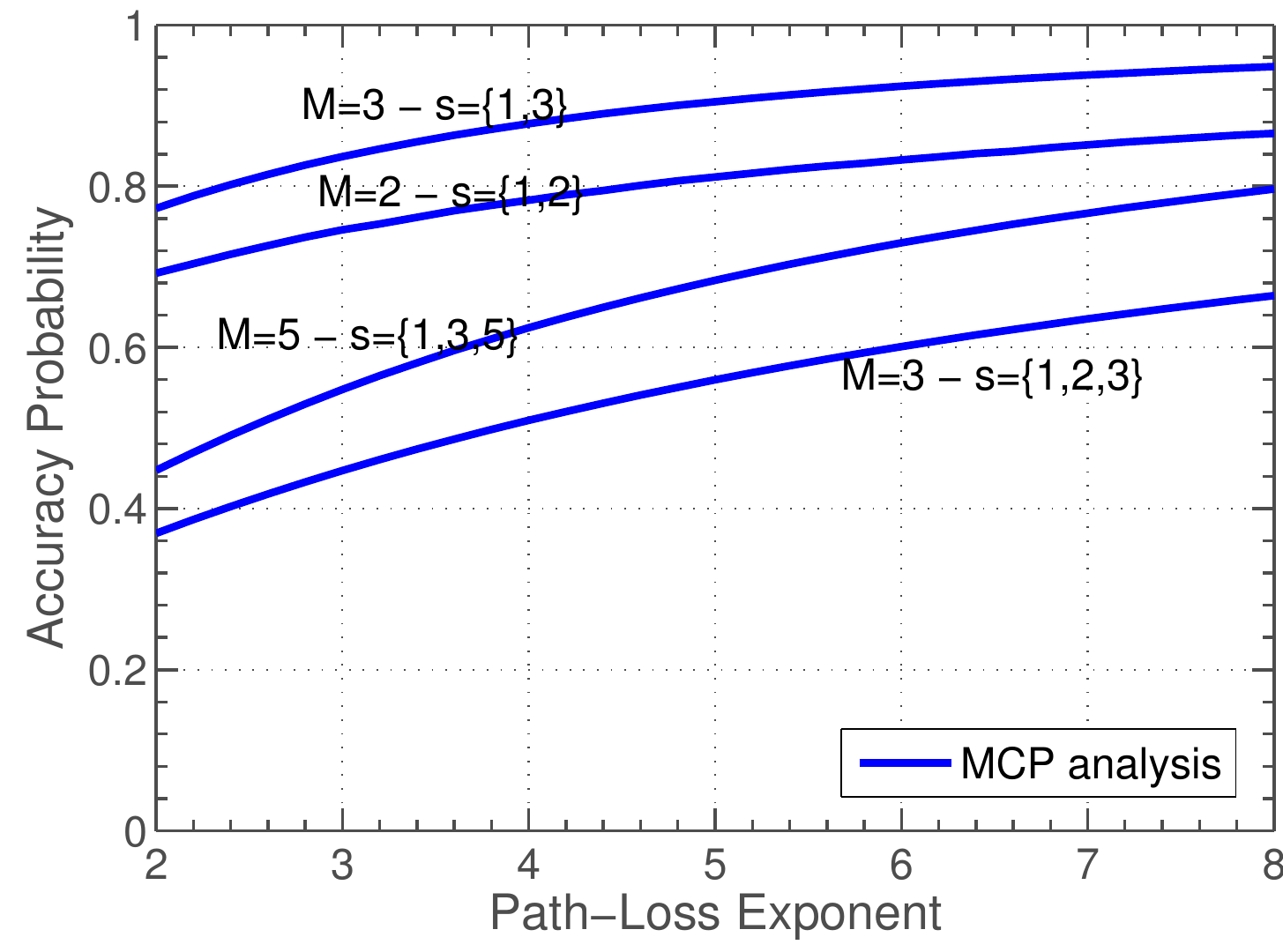,width=.49\textwidth,height=0.9\figwidth}}}} 
	\caption{Accuracy probability as a function of path-loss exponent and selecting different users for NOMA transmission with Rayleigh fading. $M$ denotes the total number of users out of which $N$ users are selected to form a NOMA cluster. Set $s$ contains ranks of the selected users for a NOMA cluster.}
	\label{fig:AvsPLE_UserPairing}
\end{figure}
The accuracy probability is shown in \figref{fig:AvsPLE_UserPairing} for 2-UE and 3-UE NOMA clusters. For instance, when two users are randomly selected, the accuracy probability for PPP with $\alpha=4$ is about 0.84. However, if we select three users randomly, and then choose the nearest and farthest users to form a NOMA cluster, the accuracy probability will be 0.92.
With more associated users with the serving BS, selection of the nearest and farthest users provide higher degree of distinctness among users. According to \figref{fig:AvsPLE_UserPairing}, with increasing channel distinctness, the accuracy probability increases significantly.

\section{Conclusion}
Most of the existing state-of-the-art analyzed NOMA performance assuming that ranking users in each NOMA cluster based on their distances, instead of the complete CSI, is a valid approximation. This approximation affects the coverage probability analysis in the uplink as well as in the downlink. This paper has verified this assumption for Rayleigh and Nakagami-$m$ fading channels and a variety of users' spatial location distributions such as PPP, MCP, and TCP. Specifically, the accuracy probability, which is the probability that the distance-based ranking matches ranking based on the instantaneous signal power, has been defined and derived. The results show that the accuracy probability is increasing with respect to the path-loss exponent while it does not depend on the BS intensity in the PPP model, cluster radius in the MCP model, and scattering variance in the TCP model. Effect of user pairing on the accuracy probability has also been investigated, and it has been shown that with distinct user pairing the accuracy probability increases significantly, compared to the random user selection.

\section*{Appendix A: Proof of Theorem~1}
\renewcommand{\theequation}{A.\arabic{equation}}
\setcounter{equation}{0}
For Rayleigh fading and $N$-UE NOMA, the inner expectation over $\{h_i\}_{i=1}^N$ can be derived as:
\begin{IEEEeqnarray}{rCl}
		\IEEEeqnarraymulticol{3}{l}{
		\mathbb{E}_{\{h_i\}}\left[ \mathbf{1} 
		\left( h_1r_{(1)}^{-\alpha}>h_2r_{(2)}^{-\alpha}>\cdots>h_Nr_{(N)}^{-\alpha} \right) \right] }
		\nonumber \\
		&=& \mathbb{E}\left[ \mathbf{1} \left( h_2r_{(2)}^{-\alpha}>\cdots>h_Nr_{(N)}^{-\alpha} \right) 
		\mathbb{E}_{h_1} \left[  \mathbf{1} \left( h_1r_{(1)}^{-\alpha}>h_2r_{(2)}^{-\alpha} \right) \right] \right]
		\nonumber \\
		&\stackrel{\text{(a)}}{=}& 
		\mathbb{E}\left[ \mathbf{1} \left( h_2r_{(2)}^{-\alpha}>\cdots>h_Nr_{(N)}^{-\alpha} \right) 
		\exp\left\{ -\frac{h_2\left(\frac{r_{(1)}}{r_{(2)}}\right)^{\alpha}}{\Omega} \right\} \right] 
		\nonumber \\
		&=& \mathbb{E}\left[ \mathbf{1} \left( h_3r_{(3)}^{-\alpha}>\cdots>h_Nr_{(N)}^{-\alpha} \right) 
		\mathbb{E}_{h_2} \left[  \mathbf{1} \left( h_2r_{(2)}^{-\alpha}>h_3r_{(3)}^{-\alpha} \right) 
		\exp\left\{ -\frac{h_2\left(\frac{r_{(1)}}{r_{(2)}}\right)^{\alpha}}{\Omega} \right\} \right] \right] \nonumber \\
		&\stackrel{\text{(b)}}{=}&
		\mathbb{E}\left[ \mathbf{1} \left( h_3r_{(3)}^{-\alpha}>\cdots>h_Nr_{(N)}^{-\alpha} \right) 
		\frac{1}{1+\left(\frac{r_{(1)}}{r_{(2)}}\right)^{\alpha}} 
		\exp\left\{ -\frac{h_3\left[ \left(\frac{r_{(1)}}{r_{(3)}}\right)^{\alpha} + \left(\frac{r_{(2)}}{r_{(3)}}\right)^{\alpha} \right]}{\Omega} \right\} \right] \nonumber \\
		&\stackrel{\text{(c)}}{=}& \frac{1}{1+\left(\frac{r_{(1)}}{r_{(2)}}\right)^{\alpha}} \cdot
		\frac{1}{1+\left(\frac{r_{(1)}}{r_{(3)}}\right)^{\alpha} + \left(\frac{r_{(2)}}{r_{(3)}}\right)^{\alpha}} \cdot \cdots \cdot
		\frac{1}{1+\left(\frac{r_{(1)}}{r_{(N)}}\right)^{\alpha} + \left(\frac{r_{(2)}}{r_{(N)}}\right)^{\alpha} \cdots  \left(\frac{r_{(N-1)}}{r_{(N)}}\right)^{\alpha}} \nonumber \\
		&=&  \prod_{i=2}^{N} \frac{1}{\sum_{j=1}^i \left(\frac{r_{(j)}}{r_{(i)}}\right)^{\alpha}},  \nonumber
	\end{IEEEeqnarray}
    where (a), (b), and (c) follow since $\left\{ h_i \right\}$ are i.i.d. exponential random variables with mean $\Omega$, the average channel power gain.

\section*{Appendix B: Proof of Theorem~2}
\renewcommand{\theequation}{B.\arabic{equation}}
\setcounter{equation}{0}
Using PDF of the link distance \eqref{eq:PDF-PPP} in \eqref{eq:final-2_UE-rayleigh} yields
	\begin{IEEEeqnarray}{rCl}
		\mathbb{E}\left[ \left(\frac{r_{(1)}}{r_{(2)}}\right)^{\alpha k} \right] 
		&=& 2 \int_{0}^{\infty} \int_{r_1}^{\infty} \left(\frac{r_1}{r_2}\right)^{\alpha k} f_{r}(r_1)f_r(r_2){\rm d}r_2{\rm d}r_1
		\nonumber \\
		&=& 2 (2c\lambda\pi)^2 \int_{0}^{\infty} \int_{r_1}^{\infty} \left(\frac{r_1}{r_2}\right)^{\alpha k} r_1 r_2 e^{-c\lambda\pi(r_1^2+r_2^2)} {\rm d}r_2{\rm d}r_1. \nonumber
	\end{IEEEeqnarray}
	Applying changes of variables $\frac{r_1}{r_2}=u$ and $r_1=v$, we have
	\begin{IEEEeqnarray}{rCl}
		\mathbb{E}\left[ \left(\frac{r_{(1)}}{r_{(2)}}\right)^{\alpha k} \right] 
		&=& 2 (2c\lambda\pi)^2 \int_{0}^{\infty} \int_{0}^{1} u^{\alpha k-3}v^3 e^{-c\lambda\pi(1+\frac{1}{u^2})v^2} {\rm d}u{\rm d}v \nonumber \\
		&=& 2 (2c\lambda\pi)^2 \int_{0}^{1} u^{\alpha k-3} \int_{0}^{\infty} v^3 e^{-c\lambda\pi(1+\frac{1}{u^2})v^2} {\rm d}v {\rm d}u. \nonumber 
	\end{IEEEeqnarray}
	Applying $c\lambda\pi(1+\frac{1}{u^2})v^2=t$ in the inner integral yields
	\begin{IEEEeqnarray}{rCl}
		\mathbb{E}\left[ \left(\frac{r_{(1)}}{r_{(2)}}\right)^{\alpha k} \right] 
		&=& 2 (2c\lambda\pi)^2 \int_{0}^{1} u^{\alpha k-3} \int_{0}^{\infty} \frac{t e^{-t}{\rm d}t} {2(c\lambda\pi)^2(1+\frac{1}{u^2})^2} {\rm d}u 
		= 4 \int_{0}^{1} \frac{u^{\alpha k+1} {\rm d}u}{(1+u^2)^2} = 2 \int_{0}^{1} \frac{x^{\alpha k/2} {\rm d}x}{(1+x)^2} \nonumber \\
		&=&\frac{2}{\frac{\alpha k}{2} + 1} \, _2F_1\left(2,\frac{\alpha k}{2} + 1;\frac{\alpha k}{2} + 2;-1\right). \nonumber
	\end{IEEEeqnarray}
	Finally, Theorem \ref{Thm2} is obtained by $_2F_1(a,b;c;z)=(1-z)^{-a}\,_2F_1(a,c-b;c;\frac{z}{z-1})$. 

\section*{Appendix C: Proof of Corollary~3}
\renewcommand{\theequation}{C.\arabic{equation}}
\setcounter{equation}{0}
From substituting \eqref{eq:PDF-PPP} in \eqref{eq:final-N_UE-rayleigh}, we get 
	\begin{IEEEeqnarray}{rCl}
		\mathcal{A}
		&=& N! (2 c \lambda \pi )^N 
		\int_{0}^{\infty} \int_{r_1}^{\infty} \cdots \int_{r_{N-1}}^{\infty} r_1 e^{-c \lambda \pi r_1^2} \prod_{i=2}^{N} 
		\frac{r_i e^{-c \lambda \pi r_i^2}} {\sum_{j=1}^i \left(\frac{r_j}{r_i}\right)^{\alpha}} {\rm d}{r_N} \cdots  {\rm d}{r_2} {\rm d}{r_1}.
		\label{eq:step1_Cor2}
	\end{IEEEeqnarray}
	Next, we simplify \eqref{eq:step1_Cor2} by applying changes of variables $r_1=u_1$, $\frac{r_{i-1}}{r_i}=u_i$ for $i=2,...,N$. Since we have $r_i=\frac{u_1}{u_2 u_3 \cdots u_i}$, $i=2,...,N$,  the Jacobian matrix $J=\frac{\partial(r_1,\cdots,r_N)}{\partial(u_1,\cdots,u_N)}$ is a triangular matrix and its determinant is equal to the product of the main diagonal entries, i.e., 
	$\det(J)=\prod_{i=1}^{N} \frac{\partial r_i}{\partial u_i} =\prod_{i=2}^{N} \frac{-u_1}{u_2 u_3 \cdots u_{i-1} u_i^2}$.
	Moreover, after changes of variables the region of integration is as $0<u_1$ and $0<u_i<1$ for $i=2,...,N$. Therefore, \eqref{eq:step1_Cor2} can be written as follows:
	\begin{IEEEeqnarray}{rCl}
		\mathcal{A}&=& N! (2 c \lambda \pi )^N 
		\int_{0}^{\infty} \int_{0}^{1} \cdots \int_{0}^{1} u_1 e^{-c \lambda \pi u_1^2} \prod_{i=2}^N 
		\left( \frac{u_1^2 u_i^{-1}\prod_{k=2}^{i}u_k^{-2}} {1+{\sum_{j=1}^{i-1} \prod_{n=j+1}^{i}u_n^{\alpha}}} 
		e^{- c \lambda \pi \frac{u_1^2}{ \prod_{m=2}^{i}u_m^{2}} } \right)
		{\rm d}{u_N} \cdots {\rm d}{u_2} {\rm d}{u_1} \nonumber \\
		&=& N! (2 c \lambda \pi )^N  \int_{0}^{1} \cdots \int_{0}^{1} 
		\prod_{i=2}^N \frac{ u_i^{-1}\prod_{k=2}^{i}u_k^{-2}} {1+{\sum_{j=1}^{i-1} \prod_{n=j+1}^{i}u_n^{\alpha}}}  \int_{0}^{\infty} 
		u_1^{2N-1} e^{-c \lambda \pi \left( 1+\sum_{i=2}^{N}\prod_{m=2}^{i}u_m^{-2} \right) u_1^2} {\rm d}{u_1} \, 
		{\rm d}{u_N}  \cdots {\rm d}{u_2}. \nonumber
	\end{IEEEeqnarray}
	Finally, \textcolor{black}{rewriting $\prod_{i=2}^N u_i^{-1}\prod_{k=2}^{i}u_k^{-2}$ as $\prod_{i=2}^N u_i^{-3-2(N-i)}$} and applying $c \lambda \pi \left( 1+\sum_{i=2}^{N}\prod_{m=2}^{i}u_m^{-2} \right) u_1^2=t$ yields
	\begin{IEEEeqnarray}{rCl}
		\mathcal{A}
		&=& N! (2 c \lambda \pi )^N  \int_{0}^{1} \cdots \int_{0}^{1} 
		\prod_{i=2}^N \frac{ u_i^{-3-2(N-i)} } {1+{\sum_{j=1}^{i-1} \prod_{n=j+1}^{i}u_n^{\alpha}}}  \int_{0}^{\infty} 
		\frac{t^{N-1}e^{-t}{\rm d}t}
		{2\left( c \lambda \pi\left[ 1+\sum_{i=2}^{N}\prod_{m=2}^{i}u_m^{-2} \right] \right)^N }\,
		{\rm d}{u_N} \cdots {\rm d}{u_2} \nonumber \\
		&=& N!(N-1)!\,2^{N-1} \int_{0}^{1} \cdots \int_{0}^{1} 
		\frac{1}{\left( 1+\sum_{i=2}^{N}\prod_{m=2}^{i}u_m^{-2} \right)^N}
		\prod_{i=2}^N \frac{ u_i^{-3-2(N-i)} } {1+{\sum_{j=1}^{i-1} \prod_{n=j+1}^{i}u_n^{\alpha}}} 
		{\rm d}{u_N} \cdots {\rm d}{u_2}, 
		\label{eq:step2_Cor2}
	\end{IEEEeqnarray}
{where the final equation is obtained by using the definition of the gamma function.} According to \eqref{eq:step2_Cor2}, $\mathcal{A}$ depends on  $\alpha$ and $N$; it does not depend on the BS intensity $\lambda$.

\section*{Appendix D: Proof of Theorem~4}
\renewcommand{\theequation}{D.\arabic{equation}}
\setcounter{equation}{0}
For Nakagami-$m$ fading and $N$-UE NOMA, the inner expectation over $\{h\}_{i=1}^N$ can be derived as:
\begin{IEEEeqnarray}{rCl}
		\IEEEeqnarraymulticol{3}{l}{
			\mathbb{E}_{\{h_i\}}\left[ \mathbf{1} 
			\left( h_1r_{(1)}^{-\alpha}>h_2r_{(2)}^{-\alpha}>\cdots>h_Nr_{(N)}^{-\alpha} \right) \right] }
		\nonumber \\
		&=& \mathbb{E}\left[ \mathbf{1} \left( h_2r_{(2)}^{-\alpha}>\cdots>h_Nr_{(N)}^{-\alpha} \right) 
		\int_{h_2\left( \frac{r_{(1)}}{r_{(2)}} \right)^{\alpha}}^{\infty}
		\frac{m^m h_1^{m-1}}{\Gamma(m) \Omega^m} e^{ -\frac{m}{\Omega} h_1 } {\rm d}h_1 \right]
		\nonumber \\
		&\stackrel{\text{(a)}}{=}& 
		\mathbb{E}\left[ \mathbf{1} \left( h_2r_{(2)}^{-\alpha}>\cdots>h_Nr_{(N)}^{-\alpha} \right) 
		\frac{m^m }{\Gamma(m) \Omega^m} h_2^{m} \left( \frac{r_{(1)}}{r_{(2)}} \right)^{\alpha m}
		\int_{1}^{\infty} t_1^{m-1} e^{-\frac{m}{\Omega}t_1 h_2\left( \frac{r_{(1)}}{r_{(2)}} \right)^{\alpha} } {\rm d} t_1 \right] 
		\nonumber \\
		&=& \mathbb{E}\Bigg[ \mathbf{1} \left( h_3r_{(3)}^{-\alpha}>\cdots>h_Nr_{(N)}^{-\alpha} \right) 
		\left( \frac{m^{m} }{\Gamma(m) \Omega^{m}} \right)^2 \left( \frac{r_{(1)}}{r_{(2)}} \right)^{\alpha m} 
		\nonumber \\
		&&\> \cdot \int_{1}^{\infty} t_1^{m-1} \int_{h_3\left( \frac{r_{(2)}}{r_{(3)}} \right)^{\alpha}}^{\infty}
		h_2^{2m-1} \exp\left\{-\frac{m}{\Omega} h_2 \left[1+t_1 \left( \frac{r_{(1)}}{r_{(2)}} \right)^{\alpha}\right]  \right\}
		{\rm d}h_2 {\rm d} t_1  \Bigg] 
		\nonumber \\
		&\stackrel{\text{(b)}}{=}& 
		\mathbb{E}\Bigg[ \mathbf{1} \left( h_3r_{(3)}^{-\alpha}>\cdots>h_Nr_{(N)}^{-\alpha} \right) 
		\left( \frac{m^{m} }{\Gamma(m) \Omega^{m}} \right)^2  \left( \frac{r_{(1)}}{r_{(2)}} \right)^{\alpha m} 
		\left( \frac{r_{(2)}}{r_{(3)}} \right)^{2 \alpha m} h_3^{2m} 
		\nonumber \\
		&&\> \cdot \int_{1}^{\infty} t_1^{m-1} \int_{1}^{\infty}
		t_2^{2m-1} \exp\left\{-\frac{m}{\Omega} h_3 \left[ t_2\left( \frac{r_{(2)}}{r_{(3)}} \right)^{\alpha}+t_1 t_2 \left( \frac{r_{(1)}}{r_{(3)}} \right)^{\alpha}\right]  \right\}
		{\rm d}t_2 {\rm d} t_1  \Bigg] 
		\nonumber \\
		&\stackrel{\text{(c)}}{=}& \left( \frac{m^{m} }{\Gamma(m) \Omega^{m}} \right)^N 
		\left( \frac{r_{(1)}}{r_{(2)}} \right)^{\alpha m} \left( \frac{r_{(2)}}{r_{(3)}} \right)^{2 \alpha m} \cdots
		\left( \frac{r_{(N-1)}}{r_{(N)}} \right)^{(N-1)\alpha m} \nonumber 
		\int_{1}^{\infty} \int_{1}^{\infty} \cdots \int_{1}^{\infty} t_1^{m-1} t_2^{2m-1} \cdots t_{N-1}^{(N-1)m-1} 
		\nonumber \\
		&&\> \cdot \int_0^{\infty} h_N^{Nm-1} \exp\left\{ -\frac{m}{\Omega} h_N
		\left[ 1+\sum_{i=1}^{N-1} \left( \frac{r_{(i)}}{r_{(N)}} \right)^{\alpha} \prod_{k=i}^{N-1}t_k \right] \right\}
		{\rm d}h_N {\rm d}t_{N-1} \cdots {\rm d}t_2 {\rm d}t_1. \nonumber
	\end{IEEEeqnarray}
	where (a), (b) are obtained by changes of variables $h_1=h_2\left( \frac{r_{(1)}}{r_{(2)}} \right)^{\alpha} t_1$, and $h_2=h_3\left( \frac{r_{(2)}}{r_{(3)}} \right)^{\alpha} t_2$. (c) follows by averaging over $h_3,...,h_N$. Finally, \textbf{Theorem \ref{Thm4}} can be obtained by applying 
	$t_N=\frac{m}{\Omega} h_N
	\left[ 1+\sum_{i=1}^{N-1} \left( \frac{r_{(i)}}{r_{(N)}} \right)^{\alpha} \prod_{k=i}^{N-1}t_k \right]$.

\section*{Appendix E: Proof of Corollary 6}
\renewcommand{\theequation}{E.\arabic{equation}}
\setcounter{equation}{0}
From \eqref{eq:match-expectation} and \textbf{Theorem \ref{Thm4}}, we can derive $\mathcal{A}$ as follows:
\begin{IEEEeqnarray}{rCl}
	\mathcal{A}&=&\frac{ \Gamma(Nm) }{ \Gamma(m)^N } 
	\int_1^{\infty} \int_1^{\infty} \cdots \int_1^{\infty} 
	\mathbb{E} \left[ 
	\frac{ \prod_{j=1}^{N-1}\left( \frac{r_{(j)}}{r_{(N)}} \right)^{\alpha m} }
	{\left[ 1+\sum_{i=1}^{N-1} \left( \frac{r_{(i)}}{r_{(N)}} \right)^{\alpha} \prod_{k=i}^{N-1}t_k \right]^{Nm}} \right]
	\prod_{j=1}^{N-1}  t_j^{jm-1} {\rm d}t_{N-1} \cdots {\rm d}t_2 {\rm d}t_1, \nonumber \\
	\label{eq:final-N_UE-Nakagami}
\end{IEEEeqnarray}
where expectation is over $\{r_{(i)}\}$. Following the same steps as \textbf{Corollary \ref{Cor2}}, we have 
\begin{IEEEeqnarray}{rCl}
	\IEEEeqnarraymulticol{3}{l}{	\mathbb{E} \left[ 
		\frac{ \prod_{j=1}^{N-1}\left( \frac{r_{(j)}}{r_{(N)}} \right)^{\alpha m} }
		{\left[ 1+\sum_{i=1}^{N-1} \left( \frac{r_{(i)}}{r_{(N)}} \right)^{\alpha} \prod_{k=i}^{N-1}t_k \right]^{Nm}} \right] =} 
	\nonumber \\
	&&\> N!(N-1)!\,2^{N-1} \int_{0}^{1} \cdots \int_{0}^{1} 
	\frac{ \prod_{j=2}^N u_j^{-3-2(N-j)+(j-1)\alpha m} }
	{ \left( 1+\sum_{i=1}^{N-1}\prod_{k=i}^{N-1}u_{k+1}^{\alpha}t_k \right)^{Nm} \left( 1+\sum_{i=2}^{N}\prod_{k=2}^{i}u_k^{-2} \right)^N}
	{\rm d}{u_N} \cdots {\rm d}{u_2}, \nonumber
\end{IEEEeqnarray}
which is independent of $\lambda$.

\section*{Appendix F: Proof of Theorem~6}
\renewcommand{\theequation}{F.\arabic{equation}}
\setcounter{equation}{0}
From \eqref{eq:match-expectation} and \eqref{eq:inner-expectation_N=2_Nakagami}, $\mathcal{A}$ can be derived as follows:
	\begin{IEEEeqnarray}{rCl}
	    \mathcal{A}&=&\frac{ \Gamma(2m) }{ \Gamma(m)^2 } 
	    \int_1^{\infty} \mathbb{E} 
	    \left[ \frac{\left( \frac{r_{(1)}}{r_{(2)}} \right)^{\alpha m}}
	    { \left[1+\left( \frac{r_{(1)}}{r_{(2)}} \right)^{\alpha} t_1 \right]^{2m}  }\right]
	    t_1^{m-1}  {\rm d}t_1
	    \label{eq:step1_Theorem6}.
	\end{IEEEeqnarray}
    In the following, we first derive the expectation in \eqref{eq:step1_Theorem6}.
    \begin{IEEEeqnarray}{rCl}
    	\mathbb{E} 
    	\left[ \frac{\left( \frac{r_{(1)}}{r_{(2)}} \right)^{\alpha m}}
    	{ \left[1+\left( \frac{r_{(1)}}{r_{(2)}} \right)^{\alpha} t_1 \right]^{2m}  }\right] &=& 2 \int_0^R \int_{r_1}^R
    	\frac{\left( \frac{r_1}{r_2} \right)^{\alpha m}}
    	{ \left[1+\left( \frac{r_1}{r_2} \right)^{\alpha} t_1 \right]^{2m}  }
    	\frac{2r_1}{R^2} \frac{2r_2}{R^2} {\rm d}r_2 {\rm d}r_1 
    	\nonumber \\
    	&\stackrel{\text{(a)}}{=}& 8 \int_1^{\infty} \int_1^{u_1} \frac{\left( \frac{u_2}{u_1} \right)^{\alpha m-3}}
    	{ \left[1+\left( \frac{u_2}{u_1} \right)^{\alpha} t_1 \right]^{2m}  } u_1^{-6} {\rm d}u_2 {\rm d}u_1  
    	\nonumber \\
    	&\stackrel{\text{(b)}}{=}& 8 \int_1^{\infty} \int_{\frac{1}{u_1}}^{1} \frac{ v^{\alpha m-3}}
    	{ \left[1+v^{\alpha} t_1 \right]^{2m}  } u_1^{-5} {\rm d}v {\rm d}u_1 
    	\nonumber \\
    	&\stackrel{\text{(c)}}{=}& 8 \int_0^{1} \int_{\frac{1}{v}}^{\infty} \frac{ v^{\alpha m-3}}
    	{ \left[1+v^{\alpha} t_1 \right]^{2m}  } u_1^{-5}  {\rm d}u_1 {\rm d}v
    	\nonumber \\
    	&=& 2 \int_0^{1} \frac{ v^{\alpha m+1}}
    	{ \left[1+v^{\alpha} t_1 \right]^{2m}  } {\rm d}v,
    	\label{eq:step2_Theorem6}
    \end{IEEEeqnarray}
    where (a) is obtained by changes of variables $r_1=\frac{R}{u_1}$ and $r_2=\frac{R}{u_2}$. (b) is obtained by applying $v=\frac{u_2}{u_1}$. (c) follows by changing the order of integrals. By substituting \eqref{eq:step2_Theorem6} in \eqref{eq:step1_Theorem6}, we get 
    \begin{IEEEeqnarray}{rCl}
    	\mathcal{A}=\frac{ 2 \Gamma(2m) }{ \Gamma(m)^2 }  \int_0^1 \int_1^{\infty}
    	\frac{ v^{\alpha m+1} t_1^{m-1} }{ \left[1+v^{\alpha} t_1 \right]^{2m}  } {\rm d}t_1 {\rm d}v. \nonumber
    \end{IEEEeqnarray}
Finally, \textbf{Theorem \ref{Thm6}} can be obtained by applying the substitution $t_1=z^{-1}$ and using the integral representation of Gaussian hypergeometric function.
    
\section*{Appendix G: Proof of Corollary~8}
\renewcommand{\theequation}{G.\arabic{equation}}
\setcounter{equation}{0}
\textcolor{black}{Assume that $M$ users are associated to the typical BS. Here we have used $[M]$ to denote the set $\{1,2,\cdots,M\}$, and we have used the notation 
	$\mathbb{P} \left( h_{\mathcal{U}_{\min}} r_{(\mathcal{U}_{\min})}^{-\alpha}> 
	                   h_{\mathcal{U}_{\max}} r_{(\mathcal{U}_{\max})}^{-\alpha} \mid \mathcal{U}=[M] \right)$ 
to emphasize that the NOMA cluster is formed by selecting the nearest and the farthest users from the set $\mathcal{U}$, where $\mathcal{U}$ includes ranks of users that we are allowed to select, $\mathcal{U}_{\min}=\min \mathcal{U}$ is the rank of the nearest user, and $\mathcal{U}_{\max}=\max \mathcal{U}$ is the rank of the farthest user. Note that removing one user from the set $[M]$ corresponds to the case that $M-1$ users are associated to the typical BS. Let assume that user $i$ is removed from the set $[M]$, and the NOMA cluster is formed by selecting the nearest and the farthest users from the set $[M]\setminus\{i\}$. Based on value of $i$, two different scenarios can occur:\\
1) When $i\in\{2,\cdots,M-1\}$, the nearest and the farthest users in the set $[M]\setminus\{i\}$ are users at rank 1 and $M$, respectively;
therefore, 	
\begin{IEEEeqnarray}{rCl}
\mathbb{P} \left( h_{\mathcal{U}_{\min}} r_{(\mathcal{U}_{\min})}^{-\alpha}>  h_{\mathcal{U}_{\max}} r_{(\mathcal{U}_{\max})}^{-\alpha} \mid \mathcal{U}=[M]\setminus\{i\} \right)=\mathbb{P} \left( h_{\mathcal{U}_{\min}} r_{(\mathcal{U}_{\min})}^{-\alpha}>  h_{\mathcal{U}_{\max}} r_{(\mathcal{U}_{\max})}^{-\alpha} \mid \mathcal{U}=[M] \right). \nonumber
\end{IEEEeqnarray}
2) When $i\in\{1,M\}$, we can show
\begin{IEEEeqnarray}{rCl}
	\mathbb{P} \left( h_{\mathcal{U}_{\min}} r_{(\mathcal{U}_{\min})}^{-\alpha}>  h_{\mathcal{U}_{\max}} r_{(\mathcal{U}_{\max})}^{-\alpha} \mid \mathcal{U}=[M]\setminus\{i\} \right)\le\mathbb{P} \left( h_{\mathcal{U}_{\min}} r_{(\mathcal{U}_{\min})}^{-\alpha}>  h_{\mathcal{U}_{\max}} r_{(\mathcal{U}_{\max})}^{-\alpha} \mid \mathcal{U}=[M] \right). \nonumber
\end{IEEEeqnarray}
Therefore, when we select the nearest and the farthest users for the NOMA cluster, the accuracy probability increases as the set of users that we are selecting from increases.
} 

\textcolor{black}{To complete the proof we need to show
\begin{IEEEeqnarray}{rCl}
	\mathbb{P} \left( h_i r_{(i)}^{-\alpha}>h_j r_{(j)}^{-\alpha} \mid [M] \right) \le \mathbb{P} \left( h_1 r_{(1)}^{-\alpha}>h_M r_{(M)}^{-\alpha} \mid [M] \right), \nonumber
\end{IEEEeqnarray}
where $i,j\in[M]$.
From \eqref{eq:match-expectation}, we have 
\begin{IEEEeqnarray}{rCl}
	\mathbb{P} \left( h_i r_{(i)}^{-\alpha}>h_j r_{(j)}^{-\alpha} \mid [M] \right) &=& 
    \mathbb{E}_{h_i,h_j} \left[ \mathbb{E}_{r_{(i)},r_{(j)}}
    \left[ \mathbf{1} \left( h_ir_{(i)}^{-\alpha}>h_jr_{(j)}^{-\alpha} \right) \right] \right] \nonumber \\
    &\stackrel{\text{(a)}}{\le}& \mathbb{E}_{h_i,h_j} \left[ \mathbb{E}_{r_{(1)},r_{(M)}}
    \left[ \mathbf{1} \left( h_ir_{(1)}^{-\alpha}>h_jr_{(M)}^{-\alpha} \right) \right] \right] \nonumber \\
    &\stackrel{\text{(b)}}{=}& \mathbb{E}_{h_1,h_M} \left[ \mathbb{E}_{r_{(1)},r_{(M)}}
    \left[ \mathbf{1} \left( h_1r_{(1)}^{-\alpha}>h_Mr_{(M)}^{-\alpha} \right) \right] \right] \nonumber \\
    &=& \mathbb{P} \left( h_1 r_{(1)}^{-\alpha}>h_M r_{(M)}^{-\alpha} \mid [M] \right), \nonumber
\end{IEEEeqnarray}
where (a) is obtained since for any realization of user point process that $h_ir_{(i)}^{-\alpha}>h_jr_{(j)}^{-\alpha}$ is satisfied, $h_ir_{(1)}^{-\alpha}>h_jr_{(M)}^{-\alpha}$ is also true, i.e., $h_ir_{(i)}^{-\alpha}>h_jr_{(j)}^{-\alpha}$ is a sufficient condition for $h_ir_{(1)}^{-\alpha}>h_jr_{(M)}^{-\alpha}$. (b) follows since all the channel power gains are i.i.d.
}

\IEEEpeerreviewmaketitle
\bibliographystyle{IEEEtran}
\bibliography{IEEEabrv,Bibliography}

\end{document}